\documentclass[prd,12pt]{article}

\usepackage{tikz}
\usepackage{amsmath,amssymb,graphicx,multirow,xspace,slashed,array,booktabs}
\usepackage[colorlinks=true,urlcolor=blue,anchorcolor=blue,citecolor=blue,filecolor=blue,linkcolor=blue,menucolor=blue,pagecolor=blue]{hyperref}
\usepackage[compress,numbers]{natbib}
\usepackage{booktabs}
\usepackage{blindtext, rotating}
\usepackage{afterpage}
\usepackage{enumitem}
\usepackage{ marvosym }
\usepackage{authblk} 
\usepackage{verbatim}
\usepackage{soul} 
\usepackage[normalem]{ulem}
\usepackage{pifont}
\usepackage{booktabs}
\usepackage{caption}
\usepackage{subcaption}
\usepackage{floatrow}
\newfloatcommand{capbtabbox}{table}[][\FBwidth]

\usepackage[font=footnotesize,labelfont=bf]{caption}

\usepackage{lineno}

\allowdisplaybreaks

\long\def\symbolfootnote[#1]#2{\begingroup%
	\def\thefootnote{\fnsymbol{footnote}}\footnote[#1]{#2}\endgroup}

\allowdisplaybreaks

\makeatletter

\@addtoreset{equation}{section}
\makeatother

\newcommand{\newc}{\newcommand}
\newc{\gsim}{\lower.7ex\hbox{$\;\stackrel{\textstyle>}{\sim}\;$}}
\newc{\lsim}{\lower.7ex\hbox{$\;\stackrel{\textstyle<}{\sim}\;$}}
\newc{\gev}{\,{\rm GeV}}
\newc{\mev}{\,{\rm MeV}}
\newc{\ev}{\,{\rm eV}}
\newc{\kev}{\,{\rm keV}}
\newc{\tev}{\,{\rm TeV}}
\newc{\MHT}{$H_T^{\text{miss}}$}
\newc{\MET}{$\slashed{E}_T$}
\newc{\MTT}{$M_{T2}$}

\def\ln{\mathop{\rm ln}}

\newc{\mz}{M_Z}
\newc{\mpl}{M_*}
\newc{\mw}{m_{\rm weak}}
\newc{\nr}[1]{N^c_R{}_{#1}}

\def\beq{\begin{equation}}
	\def\eeq{\end{equation}}
\newcommand{\bea}{\begin{eqnarray}\begin{aligned}}
		\newcommand{\eea}{\end{aligned}\end{eqnarray}}
\def\bitem{\begin{itemize}}
	\def\eitem{\end{itemize}}

\DeclareMathOperator{\sign}{sgn}
\def\vst{V_{\ast}}
\def\rst{\rho_{\ast}}
\def\pst{p_{\ast}}
\def\Omod{\Omega_{0 \Delta}}
\def\Tbrane{T_{A}^{B,\hspace{0.5mm}\rm branes} }
\DeclareFontFamily{U}{mathx}{}
\DeclareFontShape{U}{mathx}{m}{n}{<-> mathx10}{}
\DeclareSymbolFont{mathx}{U}{mathx}{m}{n}
\DeclareMathAccent{\widehat}{0}{mathx}{"70}
\DeclareMathAccent{\widecheck}{0}{mathx}{"71}

\begin{document}

\baselineskip 0.6cm

\begin{titlepage}
	
	\vspace*{-0.5cm}

	\thispagestyle{empty}

	\begin{center}
		
		\vskip 1cm

		{\Huge \bf
			Multi-brane cosmology
		}

		\vskip 1cm
		
		\vskip 1.0cm
		{\large Sudhakantha Girmohanta$^{1,2}$, Seung J. Lee$^{3}$, \\[1ex]
		Yuichiro Nakai$^{1,2}$ and Motoo Suzuki$^{4,5}$}
		\vskip 1.0cm
		{\it
			$^1$Tsung-Dao Lee Institute, Shanghai Jiao Tong University, \\
			520 Shengrong Road,
			Shanghai 201210, China \\
$^2$School of Physics and Astronomy, Shanghai Jiao Tong University, \\
800 Dongchuan Road, Shanghai 200240, China \\
		    $^3$Department of Physics, Korea University, Seoul 136-713, Korea\\
		    $^4$Department of Physics, Harvard University, Cambridge, MA 02138, U.S.A.\\
	        $^5$Institute of Particle and Nuclear Studies,\\
	        High Energy Accelerator Research Organization (KEK), Tsukuba 305-0801, Japan
    }
		\vskip 1.0cm
		
	\end{center}
	
	\vskip 1cm

	\begin{abstract}
		5D warped extra dimension models with multiple 3-branes can naturally realize multiple hierarchical mass scales
		which are ubiquitous in physics beyond the Standard Model. 
		We discuss cosmological consequences of such multi-brane models with stabilized radions.
		It is confirmed that 
		for temperatures below the scale of the IR brane at the end of the extra dimension,
		we recover the ordinary expansion of the Universe,
		with the Hubble expansion rate determined by
		sum of the physical energy densities on all 3-branes where they are localized.
		In addition, we
		explore the cosmology for temperatures above the scales of the intermediate and IR branes
		where the Universe is described by a 
        spacetime
		with the 3-branes replaced by an event horizon.
		As the temperature of the Universe cools down, phase transitions
		 are expected to take place, and the  intermediate and IR branes come out from behind the event horizon.
{The Goldberger-Wise mechanism for radion stabilization has a well-known problem of
having a supercooled phase transition,
which typically does not get completed in time. This problem is even more severe when an intermediate brane is introduced, whose scale is well above TeV, as the corresponding Hubble rate is much larger.
		 We circumvent the problem by employing an alternative mechanism for radion stabilization
		 with dark Yang-Mills fields,
		 which prevents a long supercooling epoch, but still allows the strong first order phase transitions.}
		 As a result, the phase transitions in our multi-brane Universe predict a stochastic gravitational wave background 
		 with a unique multi-peak signature,
		 which is within the sensitivity reach of future space-based gravitational wave observers.
		 {We also show that there are $N-1$ radions for an $N$ 3-brane set-up, unlike a recent claim that there exists only one radion.}
	\end{abstract}

	\flushbottom
	
	\tableofcontents
\end{titlepage}

\section{Introduction}\label{intro}
	
Theories of physics beyond the Standard Model (SM) often contain new energy scales hierarchically different from
the electroweak and Planck scales.
Such multiple hierarchical energy scales are naturally realized by introducing new 3-branes with positive tensions
into the 5D Randall-Sundrum (RS) spacetime~\cite{Randall:1999ee}
bounded by two 3-branes with positive and negative tensions called UV and IR branes.
3-branes have their typical energy scales exponentially different from each other. 
Warped extra dimension models with multiple 3-branes and their phenomenological applications have been discussed in refs.~\cite{Lykken:1999nb,Hatanaka:1999ac,Kogan:1999wc,Oda:1999di,Oda:1999be,Dvali:2000ha,Gregory:2000jc,Kogan:2000cv,Mouslopoulos:2000er,Kogan:2000xc,Pilo:2000et,Choudhury:2000wc,Kogan:2001qx,Mouslopoulos:2001uc,Kogan:2001wp,Moreau:2004qe,Agashe:2016rle,Agashe:2016kfr,Csaki:2016kqr,Fichet:2019owx,Fuentes-Martin:2020pww,Cai:2021nmk,Lee:2021slp,Fuentes-Martin:2022xnb,Girmohanta:2022giy}.
According to the AdS/CFT correspondence~\cite{Maldacena:1997re,Gubser:1998bc,Witten:1998qj},
the RS model is a holographically dual description of a nearly-conformal strongly-coupled 4D field theory
\cite{Arkani-Hamed:2000ijo,Rattazzi:2000hs}.
Then, the introduction of a new 3-brane corresponds to an extra spontaneous breaking of the conformal symmetry
via confinement in the dual 4D picture, $i.e.$ after the first spontaneous breaking of conformal symmetry, the theory flows into a new conformal fixed point~\cite{Agashe:2016rle}.
The conformal symmetry is spontaneously broken again at the IR scale, and the theory presents two distinct phase transitions.  
A multi-brane model contains multiple {\it radions}
because each distance between two 3-branes corresponds to a modulus field.
The stabilization of all the radions has been recently established in ref.~\cite{Lee:2021wau}
by a simple extension of the Goldberger-Wise (GW) mechanism
\cite{Goldberger:1999uk} introducing a bulk scalar field with brane-localized potentials.%
\footnote{However, recently there has been a claim that only one radion exists even in the presence of multiple 3-branes~\cite{Cai:2021mrw,Cai:2022geu}.
In appendix~\ref{two_radions}, we clarify important differences between ref.~\cite{Lee:2021wau} and refs.~\cite{Cai:2021mrw,Cai:2022geu} and emphasize the validity of the discussion in ref.~\cite{Lee:2021wau}.
}

If a multi-brane model is realized in nature, it must predict a consistent cosmological history of the Universe.
In the case of the RS model with two 3-branes,
there was a question of whether the late cosmology from the Big Bang Nucleosynthesis (BBN) to the present
deviates from the ordinary Friedmann–Lemaître–Robertson–Walker (FLRW) Universe.
The question has been settled in refs.~\cite{Csaki:1999mp, Csaki:2000zn}.
If the radion is stabilized correctly, the conventional FLRW equations can be recovered
in the 4D effective theory of the RS model for temperatures below the scale of the IR (TeV) brane.
It is then natural to ask if a multi-brane model with stabilized radions can reach the same conclusion,
which is discussed in the first part of the present paper.
We will confirm that for temperatures below the scale of the IR brane
the cosmology of a multi-brane model
can reproduce the ordinary FLRW Universe, with the Hubble expansion rate given by
the sum of physical energy densities on all the 3-branes. 
	
For temperatures above the typical scale of the IR brane in a multi-brane model, how does the Universe evolve?
It has been known in the RS model that the system at high temperatures is described by
an AdS-Schwarzschild (AdS-S) spacetime where the IR brane is replaced by an event horizon
\cite{Creminelli:2001th}.
As the temperature of the Universe cools down,
a phase transition from the AdS-S spacetime to the RS spacetime is expected to take place.
This phenomenon corresponds to a transition from the deconfined phase to the confined phase
in the dual 4D CFT picture.
Due to the scale invariant nature, the phase transition generally proceeds through a supercooling phase.
Ref.~\cite{Creminelli:2001th} has pointed out
that the RS model with the radion stabilized by the GW mechanism suffers from a long supercooling phase
which prevents the transition from being completed when a classical treatment of gravity is reliable
and also a backreaction effect on the gravitational action
from brane-localized potentials of the GW scalar field is negligible.
{This problem has been analyzed in detail recently~\cite{Levi:2022bzt}, and there are several interesting discussions and possible solutions~\cite{Randall:2006py,Kaplan:2006yi,Agashe:2020lfz,Agashe:2019lhy,vonHarling:2017yew,Bruggisser:2022rdm,Baldes:2021aph,Bruggisser:2018mrt,Bruggisser:2018mus,Konstandin:2011dr,Baratella:2018pxi,Fujikura:2019oyi,Csaki:2023pwy}.}
The resulting phase transition predicts the production of a stochastic gravitational wave background probed by future space-based gravitational wave observers such as LISA~\cite{LISA:2017pwj}, DECIGO~\cite{Kawamura:2020pcg}, BBO~\cite{Harry:2006fi}, TianQin \cite{TianQin:2015yph,TianQin:2020hid} and Taiji \cite{Hu:2017mde}.

In a multi-brane Universe, it is natural to expect that phase transitions take place
and intermediate 3-branes and the IR (TeV) brane come out from behind the event horizon as the temperature cools down
(see Fig.~\ref{fig:flow_diagram}).
For each phase transition to proceed, the bubble nucleation rate has to compete with
the Hubble rate at the temperature during the phase transition.
Since an intermediate 3-brane has a characteristic energy scale higher than the TeV scale
and hence the Universe has a larger Hubble rate during the phase transition,
the long supercooling problem becomes more severe. To circumvent the problem, in the present paper, we generalize the radion stabilization mechanism proposed in ref.~\cite{Fujikura:2019oyi} to a multi-brane model by introducing a dark Yang-Mills gauge field in each subregion of the bulk between two 3-branes.
Such a new radion stabilization mechanism makes it possible to avoid a long supercooling epoch,
{and still allows the strong first order phase transitions.} 
Since each phase transition associated with a 3-brane coming out from behind the event horizon produces
a stochastic gravitational wave background with some peak frequency, which is not diluted away by subsequent phase transitions,
our multi-brane model predicts a unique multi-peak gravitational wave signature detectable at future space-based observers.

The rest of the paper is organized as follows.
In section~\ref{rs3_section}, we demonstrate that one recovers the standard Hubble expansion law of the Universe
for a multi-brane model with stabilized radions.
Section~\ref{phase_transition_section} discusses phase transitions in a multi-brane Universe
where 3-branes come out from behind the event horizon.
In this section, we focus on the GW mechanism to stabilize radions
and show that the bubble nucleation rate for a phase transition associated with an intermediate brane is too small
for the transition to get completed.
Then, in section~\ref{dark_qcd_stabilization}, we consider a new radion stabilization mechanism with dark gauge fields
to address the issue.
Section~\ref{gw_section} then illustrates a unique gravitational wave signature produced during the phase transitions
in our multi-brane model.
Finally, we give conclusions and discussions in section~\ref{conclusion}.
Appendix~\ref{two_radions} emphasizes the existence of two radions in the three 3-brane setup while
some author claims that there is only one radion.

%
%
	\section{Late cosmology
	}
	\label{rs3_section}
	  We consider a 5D warped geometry with three 3-branes and study its impact on the late cosmology
	  from the BBN to the present.
	  The discussion is a generalization of that of the original RS model with two 3-branes
	  \cite{Csaki:1999mp, Csaki:2000zn}.
	  We first present our framework and then analyze conditions under which the late cosmology can reproduce the usual FLRW Universe
	   in the presence of a radion stabilization mechanism. In particular, we study conditions under which the standard Hubble law is recovered.

\subsection{The framework}
\label{framework_section}

Our spacetime geometry is given by $\mathbb{R}^4 \times S_1 / \mathbb{Z}_2$ with a background metric
\cite{Csaki:1999mp,Binetruy:1999ut},
\begin{equation}
	ds^2 = g_{AB} dx^A dx^B = n^2(t, y) dt^2 - a^2(t,y) \delta_{ij} dx^i dx^j-b^2(t,y) dy^2 \ ,
	\label{rs3_ds2}
\end{equation}
where $A,B = 0,i,4$ with $i=1,2,3$ and $x^4 \equiv y$, $\delta_{ij}$ denotes the Kronecker delta,
$y \in [0, 1/2]$ is the coordinate of the $S_1 / \mathbb{Z}_2$ orbifold,
and $n(t,y)$, $a(t,y)$ and $b(t,y)$ are all functions of the time coordinate $t$ and $y$.
They are determined by the Einstein equation,
\begin{align}
	G_{AB} = \kappa^2 T_{AB} \ . \label{Einstein_eq}
\end{align}
Here, 
$G_{AB}$ denotes the Einstein tensor, $T_{AB}$ is the energy-momentum tensor,
and $\kappa^2 \equiv 1/{(2 M_5^3)}$ with the 5D Planck scale $M_5$.  
With the metric~\eqref{rs3_ds2}, nonzero components of $G_{AB}$ are given by~\cite{Binetruy:1999ut}
\begin{align}
	\nonumber
	&G_{00}=3\left[\left(\frac{\dot{a}}{a}\right)^2+\frac{\dot{a}\dot{b}}{ab}-\frac{n^2}{b^2}\left(\frac{a''}{a}+\left(\frac{a'}{a}\right)^2\right)-\frac{a'b'}{ab}\right]\ ,\\
	\nonumber
	&G_{ii}=\frac{a^2}{b^2}\left[\left(\frac{a'}{a}\right)^2+2\frac{a'}{a}\frac{n'}{n}-\frac{b'n'}{bn}-2\frac{b'a'}{ba}+2\frac{a''}{a}+\frac{n''}{n}
	\right] \\
	\nonumber
	&\quad \quad +\frac{a^2}{n^2}\left[-\left(\frac{\dot{a}}{a}\right)^2+2\frac{\dot{a}}{a}\frac{\dot{n}}{n}-2\frac{\ddot{a}}{a}+\frac{\dot{b}}{b}\left(-2\frac{\dot{a}}{a}+\frac{\dot{n}}{n}-\frac{\ddot{b}}{b}\right)\right]\ ,\\
	\nonumber
	&G_{04}=3\left[\frac{n'}{n}\frac{\dot{a}}{a}+\frac{a'}{a}\frac{\dot{b}}{b}-\frac{\dot{a}'}{a}\right]\ ,\\
	&G_{44}=3\left[\frac{a'}{a}\left(\frac{a'}{a}+\frac{n'}{n}\right)-\frac{b^2}{n^2}\left(\frac{\dot{a}}{a}\left(\frac{\dot{a}}{a}-\frac{\dot{n}}{n}\right)+\frac{\ddot{a}}{a}\right)\right]\ ,
	\label{einstein_tensor_eqs}
\end{align}
where the prime (overdot) denotes a derivative with respect to $y$ ($t$).
We introduce UV and IR branes which extend over $\mathbb{R}^4$
and reside on the two orbifold fixed points at $y=0,1/2$, respectively.
In our setup, another intermediate 3-brane is placed at $y = y_{\rm I}$
with $0 \leq y_{\rm I} \leq 1/2$.
The UV, intermediate and IR branes have tensions, $\vst, V_{\rm I}$ and $V_{1/2}$, respectively.
The energy-momentum tensor on the branes is then given by
	\begin{align}
		\nonumber
		\Tbrane = &\frac{1}{b} \Big[ 
		\delta (y) \  {\rm diag} \big( \vst + \rst, \vst-\pst, \vst-\pst, \vst-\pst, 0  \big) \\ 
		\nonumber
		& +  \delta (y-y_{\rm I}) \  {\rm diag} \big( V_{\rm I} + \rho_{\rm I}, V_{\rm I}-p_{\rm I},  V_{\rm I}-p_{\rm I},  V_{\rm I}-p_{\rm I}, 0  \big) \\
		& +   \delta (y-1/2) \  {\rm diag} \big( V_{1/2} + \rho_{1/2}, V_{1/2} -p_{1/2},  V_{1/2} -p_{1/2}, V_{1/2} -p_{1/2}, 0  \big)
		\Big] \ .
		\label{rs3_Tab}
	\end{align}
Here, $\rst , \pst$ denote the energy density and pressure of matter localized on the UV brane,
and $\rho_{\rm I}, p_{\rm I}$ and $\rho_{1/2}, p_{1/2}$ are defined similarly.
Due to delta-functions in Eq.~(\ref{rs3_Tab}), the Einstein equation~\eqref{Einstein_eq} is solved
by decomposing it into singular and non-singular parts.
The functions $a'', n''$ will be singular,
and $a', n'$ will have finite jumps at $y=0, y_{\rm I},  1/2$.
We can relate them to the localized energy densities as jump conditions. 
We call a bulk subregion surrounded by the UV and intermediate (intermediate and IR) branes as subregion 1 (subregion 2)
and assume bulk cosmological constants, $T_{AB}^{\, \rm{subregion \, 1,2}} =  \Lambda^5_{1,2} \, g_{AB}$,
for the subregions $1,2$.

Let us now look at a static background solution for the current multi-brane setup,
taking the limit of $\rst, \rho_{\rm I}, \rho_{1/2}, \pst, p_{\rm I}, p_{1/2} \rightarrow 0$.
We begin with the following ansatz for the metric~\eqref{rs3_ds2}:
\begin{equation}
	a(t, y) = n(t, y) =  \begin{cases} 
		e^{-k_1 b_0 |y|} & 0\leq y \leq y_{\rm I} \  (\rm subregion  \ 1) \\
		e^{-k_2 b_0 |y-y_{\rm I}| - k_1 b_0 y_{\rm I}} & y_{\rm I} \leq y \leq 1/2 \  (\rm subregion  \ 2)
	\end{cases}  \  \ ; \quad b(t, y) = b_0 \ ,
	\label{rs3_static_ansatz}
\end{equation}
with curvature scales $k_{1, 2}$, and some constant $b_0$ setting the compactification radius of the extra dimension prior to orbifolding.
We can see that $a(t,y)$, or $a(y)$, is a continuous function with kinks at the branes.
Note that there are two radion modes associated with two intervals of the subregions $1,2$, which will be considered in more details in section~\ref{dark_qcd_stabilization}.
For the purpose of studying the late cosmology, the current parametrization of the background metric suffices.
The jump condition for $a(y)$ at each 3-brane is then obtained from the `$00$' component of the Einstein equation,
	\begin{equation}
		\frac{\big[ a'(y) \big]}{a(y) b_0}\bigg|_{y=0, y_{\rm I}, 1/2} = -\frac{\kappa^2}{3} V_{\rm \ast, I, 1/2} \ ,
		\label{rs3_jump_static}
	\end{equation}
where we define the jump in $f$ at a point $y$ as
\begin{equation}
	\big[f \big]\Big|_{y} = \lim_{\epsilon \to 0^+} \Big( f(y)\Big|_{y+ \epsilon} - f(y)\Big|_{y- \epsilon} \Big) \ .
	\label{deltaap_jump_def}
\end{equation}
By using the ansatz \eqref{rs3_static_ansatz}, Eq.~\eqref{rs3_jump_static} leads to
	 	\begin{equation}
	 		\vst = \frac{6 k_1}{\kappa^2} \ , \qquad
			V_{\rm I}  = \frac{3 \Delta k_{21}}{\kappa^2} \ , \qquad 
			V_{1/2} = -\frac{6 k_2}{\kappa^2} \ .
	 		\label{vst_rs3}
	 	\end{equation} 
Here, we have introduced $\Delta k_{21} \equiv k_2-k_1$.
It was noted in ref.~\cite{Lee:2021wau} that $\Delta k_{21}>0$ is required for radion stabilization.
Solving the Einstein equation for the subregions $1,2$, we obtain
 		\begin{equation}
 			\Lambda_1^5 = - \frac{6 k_1^2}{\kappa^2} \ , \qquad  \Lambda_2^5 = - \frac{6 k_2^2}{\kappa^2} \ .
 			\label{lambda_rs3}
 		\end{equation}
Eqs.~\eqref{vst_rs3}, \eqref{lambda_rs3} relate the brane tensions to the bulk cosmological constants.
Without any radion stabilization mechanism,
they require two additional fine-tuning conditions aside from the usual tuning
for a vanishing four-dimensional cosmological constant.
Once radions are stabilized, these extra fine-tuning conditions are removed.\footnote{An analogy to describe the situation is given by two positively charged balls balancing their electrical repulsion with the gravitational attraction. The static solution in this case appears only when there is an exact balance between these two forces, which requires a tuning for the charges.
However, if the two balls are connected by a spring with a sufficiently large spring constant,
the length of the spring is stabilized in its equilibrium value so that 
the fine-tuning is absorbed. 
In the present setup, a radion stabilization mechanism plays the role of such a spring.}
For a detailed discussion on the stabilization of radions based on the GW mechanism, see ref.~\cite{Lee:2021wau}. Here, we just assume potentials for the radions generated as a result of some radion stabilization mechanism. The potentials stabilize the compactification radius to $b=b_0$.

\subsection{Getting back the ordinary Universe}

In section~\ref{framework_section}, we have summarized the static solution for the three 3-brane construction.
In this limit, the background metric was given by Eq.~(\ref{rs3_static_ansatz}).
Conditions in Eqs.~\eqref{vst_rs3}, \eqref{lambda_rs3} are required to be satisfied to maintain this static solution.
However, one generally expects a deviation from this static solution
in the presence of matter, radiation and vacuum energy.
A new solution for the background metric becomes time dependent,
and it is not guaranteed that one can recover the standard FLRW Universe and preserve the success of the standard cosmology.  
To answer this question, we will perform a series expansion of the metric,
where Eq.~\eqref{rs3_static_ansatz} with Eq.~(\ref{vst_rs3}), \eqref{lambda_rs3} is considered as the zero-th order solution. 

Let us introduce matter for each brane and consider the energy-momentum tensor on the branes
given by Eq.~\eqref{rs3_Tab}. 
We denote the energy densities and pressures on the UV, intermediate and IR branes as
$\rho_{\ast, \rm I, 1/2}$, and $p_{\ast, \rm I, 1/2}$, respectively.
In the presence of these perturbations, the jump conditions in Eq.~(\ref{rs3_jump_static}) are altered.
In particular, the energy densities $\rho_{\ast, \rm I, 1/2}$ modify the jump conditions for $a(t,y)$,
and the pressures $p_{\ast, \rm I, 1/2}$ appear in the jump conditions for $n(t,y)$.
Without a potential to stabilize radions, the system of the Einstein equation together with these jump conditions appears to be over-constrained.
As discussed in ref.~\cite{Csaki:1999mp} for the two 3-brane setup,
this is the result of demanding the static solution of Eq.~\eqref{rs3_static_ansatz} with Eq.~(\ref{vst_rs3}), \eqref{lambda_rs3}, even in the absence of the radion potential.
By taking account of the radion background configurations and the solution of
the `$44$' component of the Einstein equation,
we can demonstrate that the effect of the matter perturbations is
to shift the intervals between the branes to their new equilibrium positions,
while we will discuss that the shifts are sufficiently small.
With the new intervals,
one looks for a solution for the other components of the Einstein equation.
Our goal is to establish a perturbative expansion of the metric with the matter perturbations
and study its impact on the late cosmology.  We proceed with the following ansatz for the metric tensor:
\begin{align}
	\nonumber
	a(t, y) &= 
 A(t)
 \tilde{a}(y) (1+\delta a (y)) \, , \\
	n(t, y) &=  \tilde{a}(y) (1+\delta n (y)) \, , \quad  \quad b (t,y)=b_0 \ ,
	\label{rs3_ansatz}
\end{align}
where $A(t)$ is a function of $t$ and
\begin{equation}
	\tilde{a} (y) = \begin{cases}
		e^{-k_1 b_0 |y|} & \rm{for \ subregion \ 1} \\
		e^{-k_2 b_0 |y-y_{\rm I}|-k_1 b_0 y_{\rm I}} & \rm{for \ subregion \ 2} \ .
		\label{def_atilde}
	\end{cases}
\end{equation}
Note that the time dependence is generated as a result of the matter perturbations.
At the leading order, the `$04$' component of the Einstein equation is solved trivially under the ansatz.
Furthermore, the effect of $\delta n (y)$ does not show up at this order, and hence $\rho_{\ast, \rm I, 1/2}$ determine
the solution of the remaining components of the Einstein equation. 

Our ansatz of Eqs.~(\ref{rs3_ansatz}), \eqref{def_atilde} warrants some further clarification.
As discussed above, initially, we need to consider shifts of the intervals between the branes. For the lighter radion,
its potential near the minimum scales as
$U(b) \sim m_{\rm radion}^5  \left( {\delta b}/{b_0} \right)^2$ where $m_{\rm radion}$ denotes the radion mass 
and $\delta b $ is the shift.
Hence, $\delta b$ scales inversely with the radion mass. The cosmological implication of $\delta b$ has been considered in detail in 
refs.~\cite{Csaki:1999mp, Csaki:2000zn}.
They concluded that the impact of $\delta b$ is negligibly small.
Concretely, analyzing the radion effective action, ref.~\cite{Csaki:1999mp} obtained
$\delta b/b_0 \propto \rho_{\rm NR}/(m_{\rm radion}^2 M_{\rm IR}^2)$
where $M_{\rm IR}$ denotes the typical scale of the IR brane and
$\rho_{\rm NR}$ is the energy density of non-relativistic matter.
For $T \simeq 10 \, \rm MeV$ during the BBN and the TeV scale radion, $|\delta b/b_0| \ll 10^{-20}$.
Therefore, we have assumed that the radion potential essentially fixes $b$ to its stable value $b_0$.
The similar discussion holds for the heavier radion associated with the location of the intermediate 3-brane.
The ansatz includes $\delta a$ which determines the deviation of the metric from the standard RS type solution
without the perturbations. 
A possible time dependence of $\delta a$ can change the Hubble rate $H \to H + \delta \dot{a} \simeq H(1+\delta a)$, because the typical time derivative scale is determined by $H$. If $|\delta a| \ll 1$,
we can drop this correction term at the leading order and justify our ansatz
consistent with the standard expansion of the Universe.

The Einstein equation for the bulk subregion $p \, (=1,2)$ gives
				\begin{align}
					\nonumber
					G_{00}^{\rm (bulk)} &= 3 \bigg[ H^2 -\frac{n^2}{b_0^2} \Big\{ \frac{a''}{a}+ \Big( \frac{a'}{a} \Big)^2 \Big\} \bigg] = \kappa^2 n^2 \Lambda^5_p  \ , \\[1ex]
					G_{ii}^{\rm (bulk)} &= -3 \bigg[ H^2 \frac{a^2}{n^2} -\frac{a^2}{b_0^2} \Big\{ \frac{a''}{a}+ \Big( \frac{a'}{a} \Big)^2 \Big\} \bigg] =- \kappa^2 a^2  \Lambda^5_p \ ,
					\label{G00_rs3}
				\end{align} 
			where $H=\dot A/A$ and $\Lambda^5_p=-\frac{6 k_p^2}{\kappa^2}$ from Eq.~(\ref{lambda_rs3}).
			Plugging the ansatz of Eqs.~(\ref{rs3_ansatz}), \eqref{def_atilde} into Eq.~(\ref{G00_rs3}),
			we obtain the equation for $\delta a$ at the leading order,
					\begin{equation}
						\delta a''(y) - 4 b_0 k_p \delta a'(y) = \frac{b_0^2 H^2}{\tilde{a}^2(y)} \ ,
						\label{deltaa_eqn_rs3}
					\end{equation}
					for the subregion $p$.
			The solution for $\delta a(y)$ for each subregion is given by
				\begin{align}
					\nonumber
					\delta a_1(y) &{}={} \frac{c_1}{4 k_1 b_0} \ \Big( e^{4 k_1 b_0 |y|}- 1 \Big) - \frac{H^2}{4 k_1^2} \ \Big( e^{2 k_1 b_0 |y|} - 1 \Big)  \ , \\[1ex]
					\delta a_2(y) &{}={} \frac{c_2}{4 k_2 b_0} \  e^{4 k_2  b_0 |y-y_{\rm I}|+ 4 k_1 b_0 y_{\rm I}} - \frac{H^2}{4 k_2^2} \ e^{2 k_2  b_0 |y-y_{\rm I}|+2  k_1 b_0 y_{\rm I}} + c'_2 \ ,
					\label{gen_sol_rs3}
				\end{align}
			where $\delta a_p(y)$ represents the solution for the subregion $p$
			and $c_1, c_2, c'_2$ are constants to be determined.
			We have also used $\delta a_1(0)=0$.

			The constant $c'_2$ can be expressed in terms of $c_1, c_2, H$ by requiring that $\delta a (y)$ be continuous across the intermediate 3-brane, namely $\delta a_1(y_{\rm I}) = \delta a_2 (y_{\rm I})$. 
To determine $c_1, c_2, H$, we consider the jump conditions in Eq.~(\ref{rs3_jump_static}),
including the matter perturbations,
				\begin{equation}
					\frac{\big[ a'(t, y)\big]}{a(t, y) b_0} \bigg|_{y=0, y_{\rm I}, 1/2} = -\frac{\kappa^2}{3} (V_{\rm \ast, I, 1/2} +  \rho_{\rm \ast, I, 1/2}) \ .
					\label{rs3_jump_eqs_deltaV}
				\end{equation}
			Using the ansatz with the relation (\ref{vst_rs3}), we obtain the conditions for $\delta a'(y)$, 
				\begin{equation}
					\big[\delta a'(y) \big]_{y=0, y_{\rm I}, 1/2} = - \frac{\kappa^2 b_0}{3} \rho_{\ast,\rm I, 1/2} \ .
					\label{deltaap_jumps_rs3}
				\end{equation}
Inserting Eq.~\eqref{gen_sol_rs3} into Eq.~(\ref{deltaap_jumps_rs3}), we obtain
				\begin{equation}
				\begin{split}
					&\big[\delta a'(y) \big]_{y=0} =  2 c_1 - \frac{H^2 b_0}{k_1} =  - \frac{\kappa^2}{3} b_0   \rst \ , \\[1ex]
					&\big[\delta a'(y) \big]_{y=y_{\rm I}} =   \frac{(c_2-c_1)}{\Omega_{\rm I}^4} + \frac{H^2 b_0}{2 \Omega_{\rm I}^2 } \frac{\Delta k_{21}}{ k_1 k_2}  =  - \frac{\kappa^2}{3} b_0  \rho_{\rm I} \ , \\[1ex]
						&\big[\delta a'(y) \big]_{y=1/2} =   \frac{2 c_2}{\Omod^4} -  \frac{H^2 b_0}{ k_2 \Omod^2 }  =   \frac{\kappa^2}{3} b_0  \rho_{1/2} \ ,
					\label{ceqs_rs3}
				\end{split}
				\end{equation}
				where we have defined the appropriate red-shift factors,
				$\Omega_{\rm I} \equiv e^{-k_1 b_0 y_{\rm I}}$ for the intermediate brane and $\Omod \equiv \Omega_0 \Omega_\Delta = e^{-k_2 b_0/2} e^{(\Delta k_{21}) b_0 y_{\rm I}}$ for the IR brane.
				Solving Eq.~(\ref{ceqs_rs3}), the Hubble rate is determined as
				\begin{equation}
						H^2 = \frac{ \Big( \rst + 2  \rho_{\rm I} \Omega_{\rm I}^4 +  \rho_{1/2}  \Omod^4 \Big)}{3 M_{\rm pl}^2}
						\equiv  \frac{\rho_{\rm physical}}{3 M_{\rm pl}^2} \ .
						\label{Hubble_rs3_eqn}
					\end{equation}
				Here, the 4-dimensional Planck mass is given by
				\cite{Lee:2021wau}
					\begin{equation}
						M_{\rm pl}^2 = \frac{1}{\kappa^2}  \bigg[\frac{1}{k_1} -\bigg( \frac{\Delta k_{21}}{k_1 k_2} \bigg)  \Omega_{\rm I}^2 - \frac{\Omod^2}{k_2} \bigg] \ .
						\label{Mpl_rs3}
					\end{equation}
				Eq.~(\ref{Hubble_rs3_eqn}) is the usual Hubble rate equation, where in the right hand side the physical energy density of the system determines the Hubble rate.\footnote{Note that the appearance of a factor of 2 for the energy density at the intermediate brane is an artifact of it being considered in both subregions and can be rescaled according to the convention one follows.} As expected, the energy density at each brane is warped down by the appropriate exponential suppression factor.
				The constants in Eq.~\eqref{gen_sol_rs3} are evaluated to be
					\begin{eqnarray}
						&&c_1 = \frac{b_0}{6 M_{\rm pl}^2 k_1 k_2} \left[ \rst \left(\Omega_{0 \Delta}^2 k_1 + \Omega_{\rm I}^2  \Delta k_{21} \right)   + 2 \rho_{\rm I}   k_2 \Omega_{\rm I}^4 + \rho_{1/2}   k_2 \Omega_{0\Delta}^4 \right]  ,
						\label{c1sol_eqn} \\[1ex]
						&&c_2 = \frac{b_0}{6 M_{\rm pl}^2 k_1 k_2} \left[ \rst  k_1 \Omega_{0\Delta}^2 + 2 \rho_{\rm I}  k_1 \Omega_{0\Delta}^2 \Omega_{\rm I}^4 + \rho_{1/2} \Omega_{0 \Delta}^4  \left( k_2-\Omega_{\rm I}^2 \Delta k_{21} \right) \right]  ,
						\label{c2sol_eqn} \\[1ex]
	    			&&c_2' = \frac{c_1}{4 k_1 b_0} \left( \frac{1}{\Omega_{\rm I}^4}-1 \right) - \frac{c_2}{4 k_2 b_0} \frac{1}{\Omega_{\rm I}^4} -\frac{H^2}{4 k_1^2} \left( \frac{1}{\Omega_{\rm I}^2}-1 \right) + \frac{H^2}{4 k_2^2} \frac{1}{\Omega_{\rm I}^2} \ .
	    			\label{c2p_eqn}
			\end{eqnarray}
	    		For the consistency of the calculation, we can check that the existing result for the two 3-brane setup \cite{Csaki:1999mp} is correctly recovered in the limit where the intermediate 3-brane is absent, namely $k_2 = k_1 \to k$.

      The deviation of the metric from the static solution is parametrized by  $\delta a$ in Eq.~(\ref{rs3_ansatz}). 
      Assuming no fine-tuning among the different perturbations, $|\delta a(y)|$ takes its maximum value at $y=1/2$,
      and hence we demand $|\delta a (1/2)| \ll 1$ to recover the standard Hubble expansion equation. 
       Using Eqs.~\eqref{gen_sol_rs3}, \eqref{Hubble_rs3_eqn}-\eqref{c2p_eqn}, we can write down the expression for the deviation of the metric at the IR brane,
			which is well-approximated by expanding in powers of the exponential suppression factors $\Omega_{0\Delta}$, $\Omega_{\rm I}$, $\Omega_{0\Delta}/\Omega_{\rm I}$,
							\begin{equation}
					\delta a (1/2) \simeq \frac{1}{24 k_1^2 k_2^2 M_{\rm pl}^2} \Bigg[-\rst  \frac{k_1^2}
    {\Omega_{0\Delta}^2 }
					+ 2 \rho_{\rm I} \left\{ k_2^2 - k_1^2 \frac{\Omega_{\rm I}^4}{\Omega_{0\Delta}^2}  \right\}
					+ \rho_{1/2}  k_1 k_2 \Bigg] \ .
					\label{deltaa_simplified}
				\end{equation}
Here, we have assumed that $k_1, k_2$ are not hierarchically different, in particular, $r_{12} \equiv k_1/k_2 \nless (\Omega_{0\Delta}/\Omega_{\rm I})^2$.
Without assuming fine-tuning among the different perturbations, we require each term in Eq.~(\ref{deltaa_simplified}) to be much smaller than unity,
					\begin{equation}
					\frac{\rst/M_{\rm pl}^4}{24 } \left(\frac{M_{\rm pl}}{M_{\rm IR}}\right)^2 
     \ll 1 \ ,
					\label{condition_deltaVst_simplified}
				\end{equation}
             \begin{equation}
						 \frac{\rho_{\rm I}/M_{\rm pl}^4 } { 12 \left(k_1/M_{\rm pl}\right)^2} \left|  1-
      \left(\frac{M_{\rm pl}}{k_1}\right)^2 \left( \frac{M_{\rm I}}{M_{\rm pl}} \right)^4 \left( \frac{M_{\rm pl}}{M_{\rm IR}} \right)^2  \right| \ll 1 \ ,
						\label{yI_bound}
		      \end{equation}
				\begin{equation}
						\frac{r_{12}}{24} \frac{\rho_{1/2}/M_{\rm pl}^4}{ \left(k_1/M_{\rm pl}\right)^2} \ll 1 \ ,
						\label{condition_deltaV_simplified}
				\end{equation}
where we have defined the typical intermediate mass scale $M_{\rm I} \equiv k_1 \Omega_{\rm I}$ and
the IR scale $M_{\rm IR} \equiv k_2 \Omega_{0\Delta}$.
    For the stability of the three 3-brane setup~\cite{Lee:2021wau}, $V_{\rm I}$ should be positive, and hence one requires $0 < r_{12} \leq 1$. Evidently, the constraint on the perturbation at the UV brane is $\rst \ll \left(M_{\rm IR} M_{\rm pl}
    \right)^2$. In a similar manner, from Eq.~(\ref{condition_deltaV_simplified}), we require the perturbation at the IR brane
    to satisfy $\rho_{1/2} \ll k_1^2 M_{\rm pl}^2/r_{12}$. When all the SM fields live on the IR brane, the condition \eqref{condition_deltaV_simplified} is satisfied
for temperatures below the typical scale of the IR brane, $T < M_{\rm IR}$. Furthermore, the conditions on $\rho_*,~\rho_{\rm I}$ in Eqs.~\eqref{condition_deltaVst_simplified}, \eqref{yI_bound} are satisfied if the total energy density is dominated by the contribution from the thermal bath of the SM sector, and $\rho_*,~\rho_{\rm I}$ give some small portions of the total energy density.
Therefore, we can conclude that the late cosmology at temperatures lower than $M_{\rm IR}$ follows the standard evolution.
As we will discuss in the next section,
the system at temperatures higher than $M_{\rm IR}$ is described by a different spacetime geometry
so that the current discussion is not applicable.

	\section{AdS-S and phase transitions}
	\label{phase_transition_section}

Let us now consider the system at high temperatures where the time is Euclidean and compactified on a circle. 
As discussed for the RS model with two 3-branes
\cite{Creminelli:2001th}, our current system at high enough temperatures
(above $M_{\rm I}$) is described by
an AdS-S spacetime with an event horizon.
The AdS-S Euclidean metric is given by
	\begin{align}
	\label{blackhole}
	    ds^2=\left(\frac{\rho^2}{L^2}-\frac{\rho_h^4/L^2}{\rho^2}\right)dt^2+\frac{d\rho^2}{\frac{\rho^2}{L^2}-\frac{\rho_h^4/L^2}{\rho^2}}+
	   \frac{\rho^2}{L^2}\sum_i dx_i^2\ ,
	\end{align}
	where $L=1/k_1$ and $\rho \ge \rho_h$ with the event horizon located at $\rho=\rho_h$.
	This metric gives a solution to the Einstein equation when 
	the time periodicity $\beta \equiv T^{-1}$ satisfies $\beta^{-1} = T_h \equiv \rho_h/(\pi L^2)$
	where $T_h$ is the Hawking temperature of the blackhole,
	otherwise we have a conical singularity at the event horizon.
	The dual 4D picture of the system with the real time is described by a thermal CFT with the temperature $T_h$.

The system at low enough temperatures (below $M_{\rm IR}$) is described by the RS spacetime
discussed in the previous section.
For the current purpose, we consider the following form of the metric including two radion degrees of freedom
$T_{1,2}(x)$~\cite{Lee:2021wau}:
\begin{equation}
	ds^2 = \begin{cases}
		e^{-2 k_1 T_1(x) |\phi|} \eta_{\mu \nu} dx^\mu dx^\nu - T_1(x)^2 d\phi^2  &  \text{for $0 \leq \phi \leq \phi_{\rm I}$} \\[1ex]
		e^{-2 k_2 T_2(x) |\phi'-\phi_{\rm I}'|-2 k_1 T_1(x)\phi_{\rm I}} \eta_{\mu \nu} dx^\mu dx^\nu - T_2(x)^2 d\phi'^2   &  \text{for $\phi_{\rm I}' \leq \phi' \leq \phi'_{\rm IR}$} \ ,
		\label{ds2_eqn}
	\end{cases}
\end{equation}
with the flat 4D metric $\eta_{\mu \nu}$.
This 5D metric is equivalent to the background metric in Eq.~(\ref{rs3_ds2})
when the radions are stabilized at the minimum.
Explicitly, they are related through $\phi = {b_0y}/{\langle T_1 \rangle}$, for $0 \leq y \leq y_{\rm I}$, and $\phi' = {b_0 y}/{\langle T_2 \rangle}$, for $y_{\rm I} \leq y \leq 1/2$.
The two 4D radion fields are defined as
\begin{align}
	\nonumber
	\mu_1 (x) &={} k_1 e^{-k_1 T_1(x) \phi_{\rm I}}\ , \\[1ex]
	\mu_2 (x) &={} k_2 e^{-k_2 T_2(x) (\phi_{\rm IR}'-\phi_{\rm I}')-k_1 T_1(x) \phi_{\rm I}}\ .
	\label{radion_defns}
\end{align}
The stabilization of all the 3-branes amounts to generating a 4D effective potential for the radions with a non-trivial minimum at $\mu_{1, \rm min}$ and $\mu_{2, \rm min}$, which give the typical energy scales at the intermediate and IR branes, respectively.
The radion fields defined in Eq.~\eqref{radion_defns} are not canonically normalized.
In fact, their kinetic terms are~\cite{Lee:2021wau} 
\begin{equation}
	S_{\rm kin, rad} = \int d^4x \left[  \frac{12 M_5^3}{k_1^3} \left(1-\frac{k_1}{k_2}\right) \eta^{\rho \sigma} \partial_{\rho} \mu_1  \partial_{\sigma} \mu_1 + \frac{12 M_5^3}{k_2^3}  \eta^{\rho \sigma} \partial_{\rho} \mu_2  \partial_{\sigma} \mu_2 \right]  \ .
	\label{radion_kin:eq}
\end{equation} 
Note that the first term requires $k_2 > k_1$. In the limit of $k_2 \rightarrow k_1$, this term becomes singular,
which signals the disappearance of the intermediate brane.
Generally, the mass of the radion $\mu_1$ is around the energy scale of the intermediate brane
and much heavier than that of $\mu_2$ determined by the scale of the IR brane.
 \begin{figure}
	\centering
	\hspace{0cm}
	\includegraphics[width=0.9\linewidth]{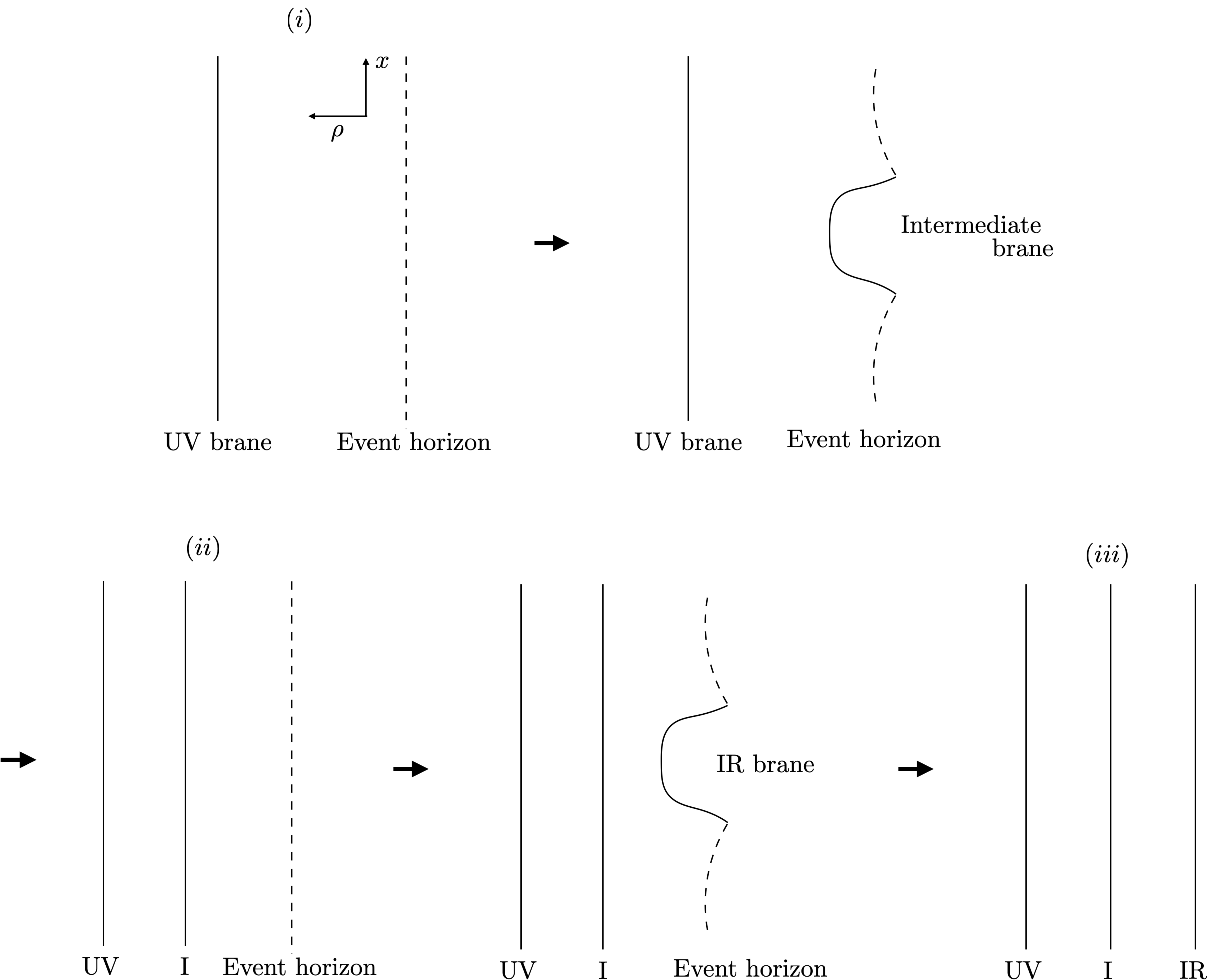}
	\vspace{0.2cm}
	\caption{A schematic diagram of phase transitions via bubble nucleation in the three 3-brane model.
	There are three regimes: (i) the AdS-S spacetime with an event horizon, (ii) the AdS-S spacetime with the intermediate 3-brane and
    (iii) the time-compactified RS spacetime with the intermediate and IR branes. 
    The first phase transition from the regime (i) to (ii) takes place through the nucleation of the intermediate brane bubbles.
    The second transition from the regime (ii) to (iii) takes place through the nucleation of the IR brane bubbles.}
	\label{fig:flow_diagram}
\end{figure}

    Since the system behaves differently at high and low temperatures,
    phase transitions take place between different regimes as the Universe cools down.
    In the current three 3-brane setup, we can consider three regimes (see Figure~\ref{fig:flow_diagram}): (i) the AdS-S spacetime with an event horizon, (ii) the AdS-S spacetime with the intermediate 3-brane and
    (iii) the time-compactified RS spacetime with the intermediate and IR branes. 
    Under the assumption of a large distance separation between the intermediate and IR branes,
    $\mu_{\rm 2, min} \ll \mu_{1, \rm min}$,
    we expect two phase transitions, $i.e.$ the first transition from the regime (i) to (ii)
    and the second transition from the regime (ii) to (iii).
    In the dual 4D picture, they correspond to two confinement-deconfinement phase transitions.
    As the Universe cools down, the thermal CFT undergoes a confinement at the first phase transition and flows into another thermal CFT,
    which is in the end confined at the second transition.
    The mixing between two dilatons is suppressed by $\sim \epsilon \left(\mu_{\rm 2, min}/\mu_{1, \rm min}\right)^3$ with a conformal symmetry breaking parameter $\epsilon$
    \cite{Agashe:2016kfr}. 
    Therefore, in the 5D picture, it is justifiable to assume that two radions are decoupled
    when $\mu_{\rm 2, min} \ll \mu_{1, \rm min}$.
    
    Let us first consider the phase transition from the regime (i) to (ii).
    As in the case of the two 3-brane model~\cite{Creminelli:2001th},
    we can expect that the phase transition proceeds through the pure AdS spacetime with the periodic time.
    That is, moving from the outside to the center of a nucleating bubble, we observe that the horizon goes toward
    $\rho = 0$ and then the intermediate brane comes in from  $\rho = 0$ to a value at the minimum. 
    To evaluate the critical temperature of the phase transition, we need to find the free energy of the system in each regime.
For the regime (i), the free energy is dominated by the black hole solution \eqref{blackhole},
and we assume that the contribution from a radion potential is sub-dominant
because of its negligible back-reaction to the background geometry.
Taking account of the conical singularity effect,
the difference between the free energy of the AdS-S spacetime and that of the pure AdS is calculated as
\cite{Creminelli:2001th}
	\begin{align}
		\nonumber
		\Delta F_{\rm regime \, (i)} &= 6 \pi^4 \left(\frac{M_5}{k_1}\right)^3 T_h^4 - 8 \pi^4 \left(\frac{M_5}{k_1}\right)^3 T T_h^3
		\\[1ex]
		&=\frac{3}{8} \pi^2 N_1^2 T_{h}^4 - \frac{1}{2} \pi^2 N_1^2 T_{h}^3 T 
		\label{F_Ads-S:eq} \ ,
	\end{align}
where we have defined
\begin{equation}
	\frac{N_1^2}{16 \pi^2} \equiv \frac{M_5^3}{k_1^3} \quad ; \qquad \frac{N_2^2}{16 \pi^2} \equiv \frac{M_5^3}{k_2^3} \ ,
\end{equation}
which roughly give the numbers of colors in the two corresponding dual CFT theories.
The free energy of Eq.~\eqref{F_Ads-S:eq}, as a function of $T_h$, has the minimum at $T_h=T$. 
For the regime (ii),
we assume that the critical temperature of the phase transition is somewhat smaller than
the typical scale of the intermediate brane so that the heavier radion $\mu_1$ is no longer in the thermal bath.
The contribution to the free energy is then approximated by a potential stabilizing the heavier radion.
Another contribution may come from the second thermal CFT with the number of colors $N_2$ in the dual 4D picture.
This effect is, however, neglected if its free energy is smaller than that of the first thermal CFT
with the number of colors  $N_1$, $i.e.$ $N_1> N_2$.
The intermediate brane is assumed not to contain a lot of light degrees of freedom
so that their contribution to the free energy can be safely neglected. In the end, the free energy for the regime (ii) subtracted by that of the pure AdS is given by
	\begin{equation}
		\Delta F_{\rm regime \, (ii)} = V_{\rm r, eff, 1} (\mu_{\rm 1, min}) - V_{\rm r, eff, 1} (0) \ .
		\label{F_RS:eq}
	\end{equation}
Here, $V_{\rm r, eff, 1} (\mu_{\rm 1})$ denotes a potential stabilizing the heavier radion.
We will discuss two different types of the potential below.
At the critical temperature $T_{\rm c, 1}$, the free energy at its minimum for the regime (i) is equal to
that for the regime (ii). By using Eqs.~\eqref{F_Ads-S:eq}, \eqref{F_RS:eq}, we find
	\begin{equation}
		T_{\rm c, 1} = \left(\frac{8 |\Delta F_{\rm regime \, (ii)}|}{\pi^2 N_1^2}\right)^{1/4} \ ,
		\label{critical_temp:eq}
	\end{equation}
which is somewhat suppressed by a large $N_1$ factor and justifies our assumption. 

    Under the assumption of a sufficiently large distance separation between the intermediate and IR branes,
    we can discuss the phase transition from the regime (ii) to (iii) similarly. 
    That is, the relevant part of the free energy for the regime (ii) is now given by the minimum of Eq.~\eqref{F_Ads-S:eq}
    with $N_1$ replaced by $N_2$,
    and the free energy for the regime (iii) comes from a potential stabilizing the lighter radion $V_{\rm r, eff, 2} (\mu_{\rm 2})$.
    The critical temperature $T_{\rm c, 2}$ is estimated accordingly.

	\subsection{Goldberger-Wise potential for radion stabilization}
	
	To discuss the phase transitions more concretely, the radion potential needs to be specified.
	In this section, we consider a single 5D scalar field $\phi$ with a bulk mass $m_\phi$ and localized brane potentials~\cite{Lee:2021wau}
	as in the case of the original GW mechanism for the two 3-brane setup~\cite{Goldberger:1999uk}.  
		The lighter radion potential is similar to that of the two 3-brane case, so we mainly focus on the potential for the heavier radion $\mu_1$.
	The low-energy 4D effective potential is obtained after integrating out the extra dimension
	(see Eq.~(A.8) in ref.~\cite{Lee:2021wau}),
		\begin{align}
			\nonumber
			V_{\rm GW}(\mu_1) =& \, \mu_1^4 \left[v_{\rm I} - v_{\rm UV} \left(\frac{\mu_1}{k_1}\right)^\epsilon\right]^2 (4+ \epsilon) - \epsilon \mu_1^4 v_{\rm UV} \left(\frac{\mu_1}{k_1}\right)^\epsilon \left[2 v_{\rm I} - v_{\rm UV} \left(\frac{\mu_1}{k_1}\right)^\epsilon \right] \\[1ex]
												=& \, \mu_1^4 \left[ v_{\rm I} - v_{\rm UV} \left(\frac{\mu_1}{k_1}\right)^\epsilon \right]^2 (4 + \epsilon) + \mu_1^4 \epsilon \left[ v_{\rm I} - v_{\rm UV} \left(\frac{\mu_1}{k_1}\right)^\epsilon \right]^2 -\mu_1^4 \epsilon v_{\rm I}^2 \ .
			\label{V1_lee_suzuki_nakai}
		\end{align}
	Here, the boundary potentials for $\phi (y)$ are $V_{\rm UV} = \lambda_{\rm UV} (\phi(0)^2 - v_{\rm UV}^2 k_1^3)^2$, $V_{\rm I} = \lambda_{\rm I} (\phi(y_{\rm I})^2 - v_{\rm I}^2 k_1^3)^2$
	with dimensionless $v_{\rm UV}, v_{\rm I}$, a constant term is subtracted from the radion potential and $\mathcal{O}(\epsilon^2)$ pieces are dropped. In addition, we have defined $\epsilon \equiv \sqrt{4 + m_\phi^2/k_1^2}-2 \simeq m_\phi^2/4k_1^2$, which is a small parameter and $\epsilon \geq -2$~\cite{Breitenlohner:1982bm}.
	We will take it to be positive for concreteness.
	Simplifying Eq.~(\ref{V1_lee_suzuki_nakai}), we arrive at
		\begin{equation}
			V_{\rm GW}(\mu_1) = \mu_1^4 v_{\rm I}^2 \bigg[ (4 + 2 \epsilon) \bigg\{1- \frac{1}{R_v} \bigg(\frac{\mu_1}{k_1} \bigg)^\epsilon \bigg\}^2 -\epsilon - \widetilde{\delta} \bigg] \ ,
			\label{radion_potential_eqn}
		\end{equation}
	where we have introduced the ratio $R_v \equiv v_{\rm I}/ v_{\rm UV}$. The last term $\propto - \widetilde{\delta} \mu_1^4 v_{\rm I}^2$ represents a deviation from the static RS solution, for instance, due to loop corrections.\footnote{
	Notice a relative sign change from ref.~\cite{vonHarling:2017yew} in the definition of $\widetilde{\delta}$}
	Given that $\widetilde{\delta}$ lies in the interval,
		\begin{equation}
			I_{\widetilde{\delta}} : -(\epsilon + \epsilon^2/4) <  \widetilde{\delta} <  4 + \epsilon  \ ,
		\end{equation}
	$V_{\rm GW}$ has a global minimum (and a maximum) at 
		\begin{align}
			\nonumber
			\mu_{\rm 1, min, max} &{}\simeq{} k_1 R_v^{1/\epsilon} X_{\rm min, max}^{1/\epsilon} \ , \\[1ex]
			X_{\rm min, max} &{}={} \frac{1}{(1+\epsilon/2)} \bigg[ 1 + \frac{\epsilon}{4} \pm \frac{\sign{\epsilon}}{2} \sqrt{\epsilon + \frac{\epsilon^2}{4}+ \widetilde{\delta}} \ \bigg] \ .
			\label{mumin_eqn}
		\end{align}
	Note that $X_{\rm min} > X_{\rm max}$, so that the global minimum of the potential is separated from the origin by a potential barrier situated at $\mu_{\rm 1, max}$.
	
	The potential $V_{\rm GW}(\mu_1)$ has a form $\mu_1^4 P(\mu_1^\epsilon)$ with a polynomial function $P$, and hence represents an almost marginal operator of conformal dimension $4+ \epsilon$ in the dual 4D picture.
	Therefore, for a small $\epsilon$, the conformal symmetry is broken through the slow renormalization group (RG) evolution,
	so that a hierarchically small mass scale $\mu_{\rm 1, min}$ is generated.
	%
	\begin{figure}[t]
		\begin{center}
			\hspace{0.3cm}
			\begin{subfigure}{0.45\textwidth}
				\centering
				\includegraphics[width=\textwidth]{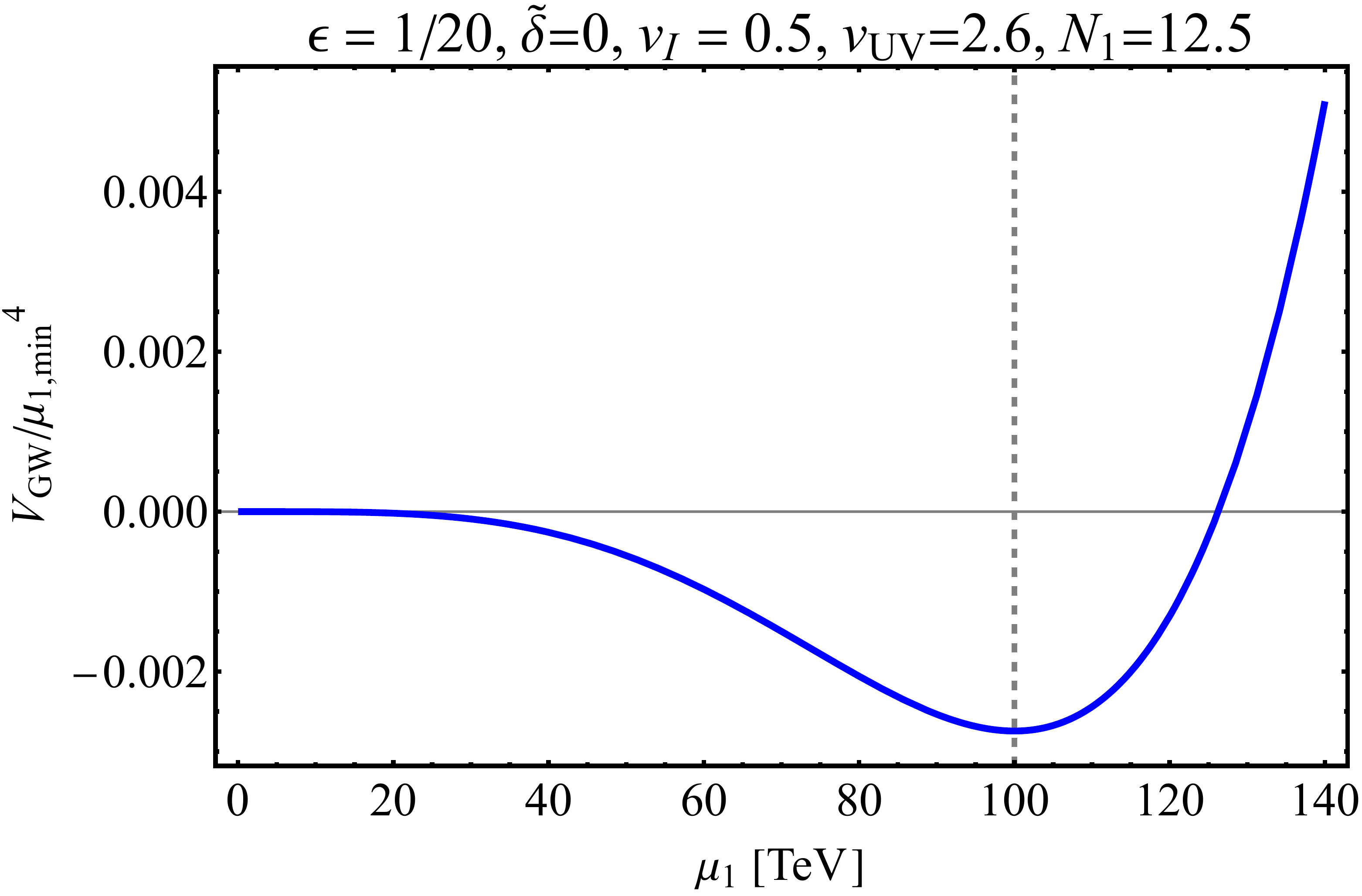}
				\subcaption{}
				\label{vgw_mumin_del0}
			\end{subfigure}
			\hfill
			\hspace{0.5cm}
			\begin{subfigure}{0.45\textwidth}
				\centering
				\includegraphics[width=\textwidth]{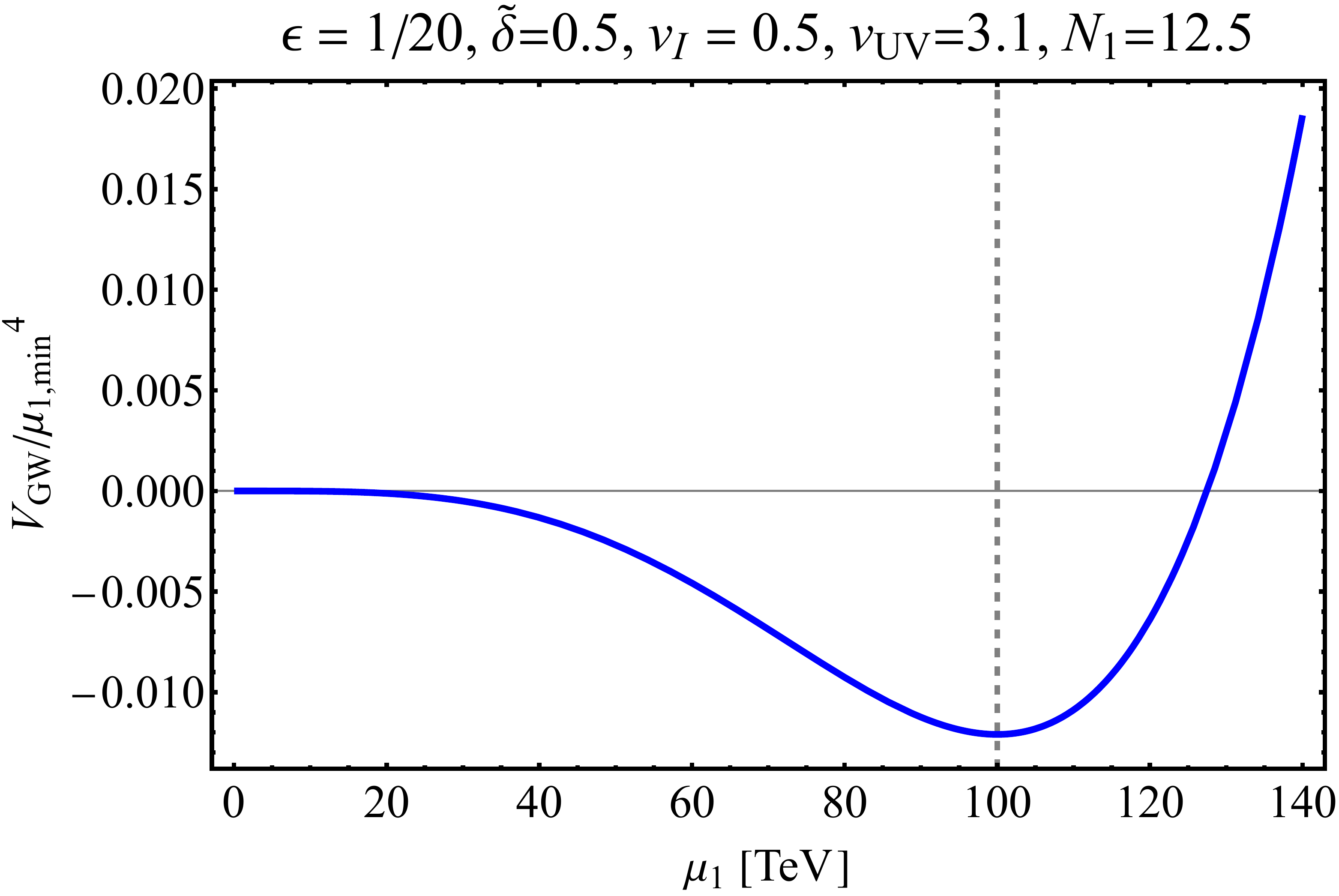}
				\subcaption{}
				\label{vgw_mumin_delnonzero}
			\end{subfigure}
			\hfill
			\vspace{0.5cm}
			\begin{subfigure}{0.48\textwidth}
				\centering
				\includegraphics[width=\textwidth]{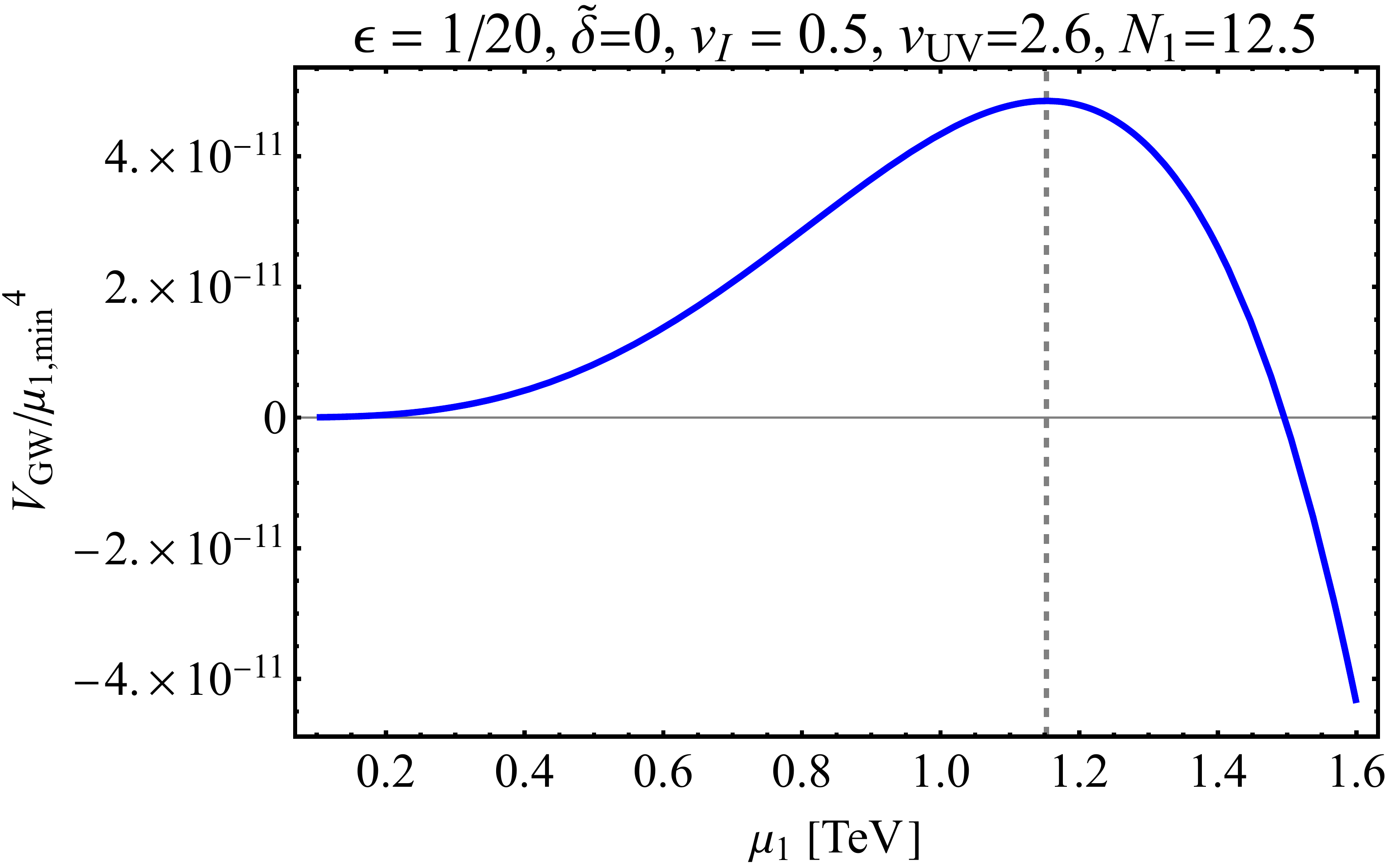}
				\subcaption{}
				\label{vgw_mumax_del0}
			\end{subfigure}
			\hfill
			\begin{subfigure}{0.48\textwidth}
				\centering
				\includegraphics[width=\textwidth]{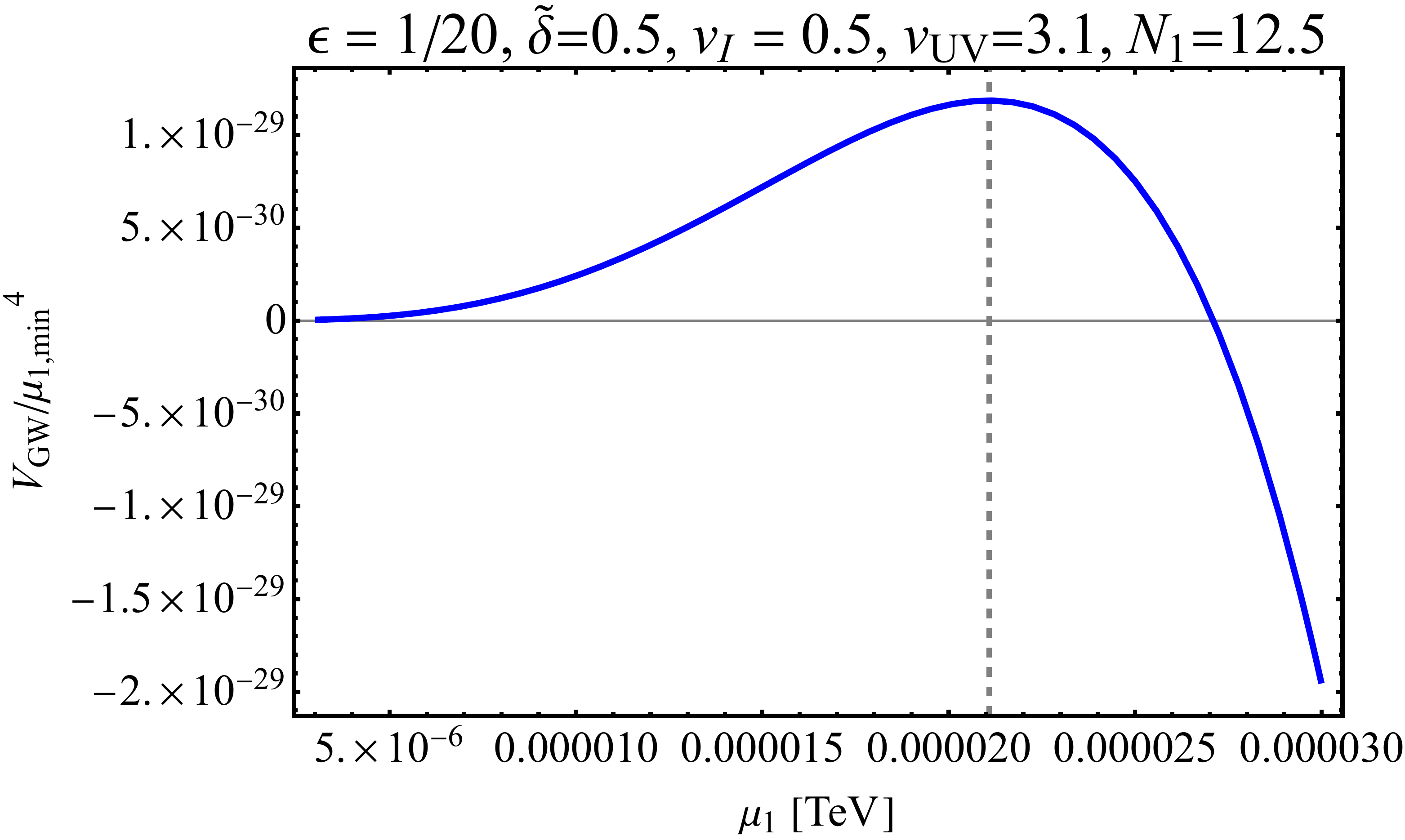}
				\subcaption{}
				\label{vgw_mumax_delnonzero}
			\end{subfigure}
		\end{center}
		\vspace{-0.3cm}
		\caption{The effect of $\widetilde{\delta}$ on the GW potential for the heavier radion.
		We take $\widetilde{\delta} = 0, 0.5$ in the left and right panels, respectively.
		The bottom panels magnify the regions abound the origin of $\mu_1$ in the top panels. 
		For $\widetilde{\delta} \in I_{\widetilde{\delta}}$ and $\widetilde{\delta} > \epsilon$,
		the  minimum of the potential gets deeper and the potential barrier is significantly diminished.} 
		\label{vgw_comp_figs}
	\end{figure}
	%
	%
	When the vacuum expectation value (VEV) of $\phi$ is large, its backreaction can deform the geometry
	and the current analysis is not reliable. To avoid this, we impose
	\cite{vonHarling:2017yew}
		\begin{equation}
			\frac{v_{\rm I}}{N_1} \ll \min\bigg[\frac{1}{2 \pi X_{\rm min}} \sqrt{\frac{{3}}{\epsilon}} \bigg(\frac{\mu_{\rm 1, min}}{k_1}\bigg)^\epsilon, \frac{\sqrt{3}}{4 \pi} \bigg] \ .
			\label{backreaction_eqn}
		\end{equation} 
	To fix $\mu_{\rm 1, min}$ in Eq. (\ref{mumin_eqn}) for a given $\epsilon$, the ratio $R_v$ has to be kept constant, which therefore implies 
		\begin{equation}
			\frac{v_{\rm UV}}{N_1} \ll \min\bigg[\frac{1}{2\pi} \sqrt{\frac{3}{\epsilon}}, \frac{\sqrt{3}}{4\pi} X_{\rm min} \left( \frac{k_1}{\mu_{\rm 1, min}} \right)^\epsilon \bigg] \ .
			\label{backreaction_eqn_vUV}
	   	\end{equation} 
   	Moreover, in order for the perturbative description of our 5D theory to be reliable,
	we assume $N_{1, 2} \gtrsim 4.4$, which ensures that higher powers of Ricci scalars from the quantum gravity effect can be safely neglected~\cite{Agashe:2007zd}.
	
	The parameter $\widetilde{\delta}$ makes the minimum of the potential deeper
	and reduces the potential barrier significantly.
	Concretely, we perform a series expansion in terms of $\epsilon$ and $\widetilde{\delta}$, and the radion potential (\ref{radion_potential_eqn}) can be written as
		\begin{equation}
			V_{\rm GW}(\mu_1) \simeq \mu_1^4 v_{\rm I}^2 \bigg[ (4+2 \epsilon) \bigg\{ 1- X_{\rm min} \bigg(\frac{\mu_1}{\mu_{\rm 1, min}}\bigg)^\epsilon \bigg\}^2- \epsilon - \widetilde{\delta} \bigg] \ .
			\label{vgw_in_mumin}
		\end{equation}
	For $\widetilde{\delta} \gg \epsilon$, at the leading order, we obtain
		\begin{equation}
			\frac{V_{\rm GW} (\mu_{\rm 1, min})}{\mu_{\rm 1, min}^4} \simeq - \frac{1}{2} v_{\rm I}^2  \hspace{0.5mm}  \epsilon  \hspace{0.5mm} {\widetilde{\delta}}  + \cdots \ ,
			\label{vmin_GW_delta:eq}
		\end{equation}
	whereas, for $\widetilde{\delta}=0$, at the leading order,
			\begin{equation}
		\frac{V_{\rm GW} (\mu_{\rm 1, min})}{\mu_{\rm 1, min}^4} \simeq - v_{\rm I}^2  \hspace{0.5mm}  \epsilon  \hspace{0.5mm} \sqrt{\epsilon}  + \cdots \ .
		\label{Vmin_GW_epsilon:eq}
	\end{equation}
Hence, as long as $\widetilde{\delta} > 2 \sqrt{\epsilon}$, the minimum of the potential gets deeper, as we show in figure~\ref{vgw_comp_figs}.
	Note that the potential remains shallow because $V_{\rm GW}(\mu_1)$ is still nearly conformal invariant
	as a scale only enters as $(\mu_1/k_1)^\epsilon$.

\subsection{Does the phase transition get completed?}

Now, let us discuss the phase transition from the regime (i) to (ii), assuming the GW potential $V_{\rm GW}(\mu_1)$.
In order for the phase transition to be completed, during the phase transition, the bubble nucleation rate should be greater than the Hubble rate per unit horizon time and volume,
	\begin{equation}
	\Gamma_n = \Gamma_0 e^{-S} \gtrsim H^4 \ ,
	\end{equation}
where the action $S$ is dominated by $O(4)$ symmetric bubbles for temperatures smaller than $\mu_{\rm 1, min}$, and $\Gamma_0$ denotes the relevant energy scale of the potential raised to the fourth power.
For our purpose of getting an estimate, the zero temperature action for the $O(4)$ symmetric bubbles suffices,
	\begin{equation}
		S_4 \simeq \frac{9 N_1^4}{8 \pi^2} \frac{\mu_{1, r}^4}{-V_{\rm GW}(\mu_{1, r})} \ .
	\end{equation}
Here, $\mu_{1, r}$ is the field value which the field tunnels to and determined by minimizing $S_4$. This can be done numerically, but for our rough estimation purpose, we can take $\mu_{1, r} \sim \mu_{\rm 1, min}$ as in ref.~\cite{vonHarling:2017yew}. The Hubble rate during the phase transition is $H \simeq \mu_{\rm 1, min}^2/M_{\rm pl}$, which is valid for temperatures near or below the critical point. Note that the larger the intermediate brane mass scale is,
the larger the Hubble rate is, and the bubble nucleation rate has to compete with this larger expansion rate for the phase transition to proceed. Therefore, in order for the phase transition to be completed, we require
	\begin{equation}
		\frac{9 N_1^4}{32 \pi^2} \frac{\mu_{\rm 1, min}^4}{-V_{\rm GW}(\mu_{\rm 1, min})^4} \lesssim \ln\bigg(\frac{M_{\rm pl}}{\mu_{\rm 1, min}}\bigg)\ ,
		\label{phase_transition_condition}
	\end{equation}
where we assume $\Gamma_0 \simeq \mu_{\rm 1, min}^4$.

It is evident that there is a tug-of-war between the back reaction condition in Eq. (\ref{backreaction_eqn}), and Eq. (\ref{phase_transition_condition}), namely that in order to satisfy the back reaction condition, we prefer a large $N_1$, but the bubble nucleation action scales as $N_1^4$, which should be small in order for the phase transition to proceed. In addition, ${-V_{\rm GW}(\mu_{\rm 1, min})}/{\mu_{\rm 1, min}^4}$ is preferred to be large, by taking $\epsilon, \delta$ as large as possible, to make the potential deep. Although $\widetilde{\delta}$ helps deepening the potential, it is bound to live in the interval $I_{\rm \widetilde{\delta}}$. Moreover, one can not make $\epsilon$ arbitrarily large, keeping $\mu_{\rm 1, min}$ fixed,  as $v_{\rm UV}$ increases rapidly and eventually goes beyond the $v_{\rm UV} \ll N_1$ regime. Now, looking at the right hand side of the inequality (\ref{phase_transition_condition}), we see that increasing $\mu_{\rm 1, min}$ makes it difficult for the phase transition to be completed. Therefore, without the help of other sources of conformal breaking, the transition from the regime (i) to (ii) does not get completed.
The numerical calculation has also confirmed this conclusion. 

\section{Radion stabilization with dark gauge fields}
\label{dark_qcd_stabilization}
\begin{figure}[!t]
	\begin{center}
			\includegraphics[width=0.4\textwidth]{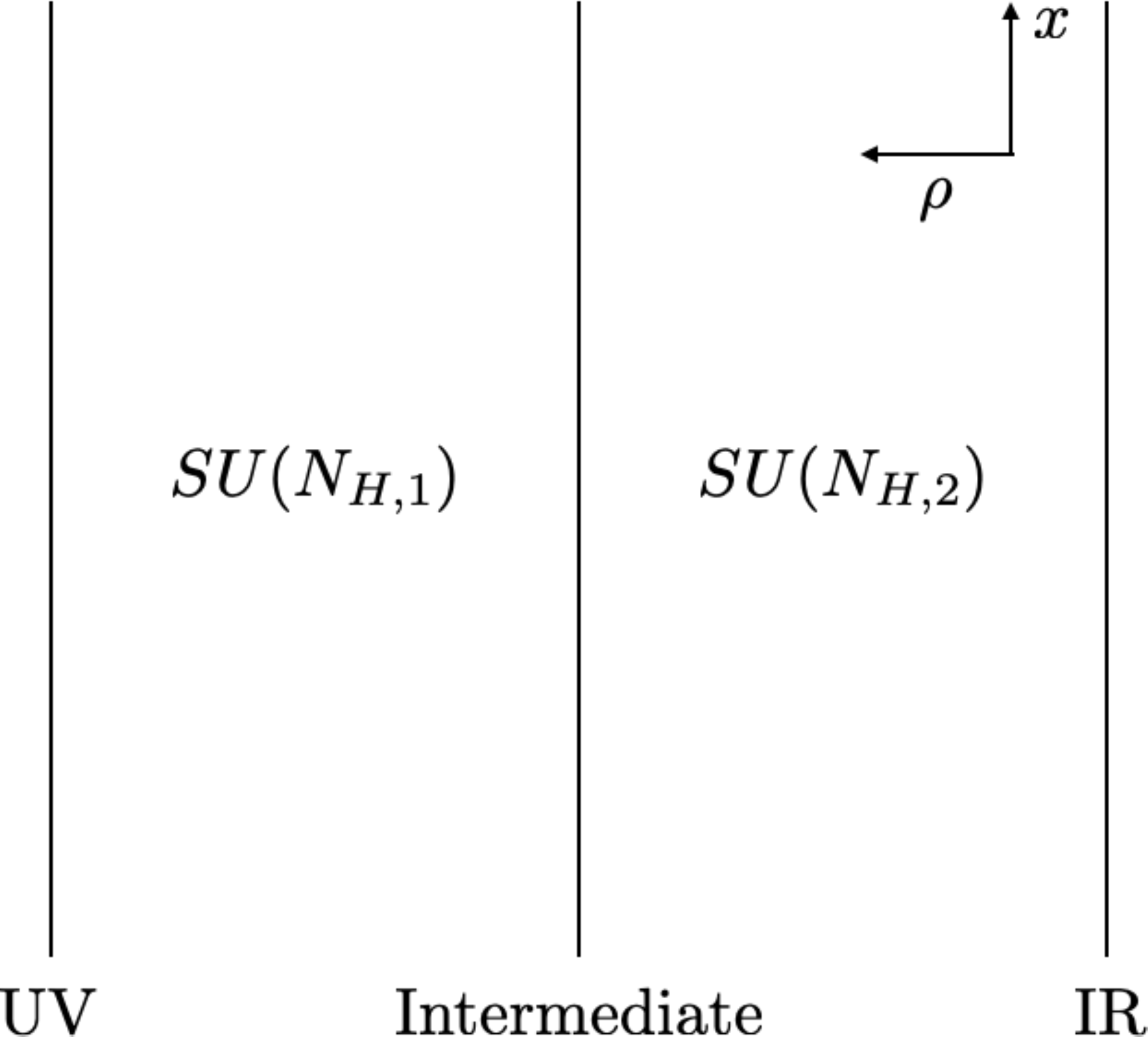}
	\end{center}
	\caption{The schematic description of our setup to stabilize radions utilizing dark gauge fields.} 
	\label{figs_dark_qcd_setup}
\end{figure}
%
%

In this section, we consider an alternative mechanism to stabilize the radions in the three 3-brane model
to make the phase transitions completed correctly.
Ref.~\cite{Fujikura:2019oyi} has presented a radion stabilization mechanism for the two 3-brane model
with a bulk confining gauge field, where
the deviation of the IR brane tension from the static condition contributes to a quartic potential for the radion,
which is balanced by a potential generated from the gauge field confinement.
We here generalize this mechanism for our setup and discuss the phase transitions
with the new radion potentials.

\subsection{The model}

Let us introduce two $SU(N_{H, i})$ gauge fields residing respectively in the subregions $i = 1, 2$, as described in figure~\ref{figs_dark_qcd_setup}.
Their 5D action with the metric defined in Eq.~(\ref{ds2_eqn}) is given by
\begin{equation}
	\begin{split}
		S &\supset S_{\rm YM, 5}^{(1)} +  S_{\rm YM, 5}^{(2)} \\[1ex]
		&= \int d^4 x \bigg[ \int_{0}^{\phi_{\rm I}} d\phi  \sqrt{G} \,  \bigg( -\frac{1}{4 g_{5, 1}^2} F_{MN, 1} F^{MN, 1}  \bigg)  
		 +  \int_{\phi_{\rm I}'}^{\phi_{\rm IR}'} d\phi'  \sqrt{G}  \, \bigg( -\frac{1}{4 g_{5, 2}^2} F_{MN, 2} F^{MN, 2}  \bigg) \bigg] \ ,
	\end{split}
	\end{equation}
where $F_{MN,i}$ and $g_{5, i}$ denote the 5D gauge field strength and coupling for the $SU(N_{H,i})$.
In addition, we consider deviations of the brane tensions from their static tuned values in Eq.~(\ref{vst_rs3}), namely 
\begin{equation}
		\vst = \frac{6 k_1}{\kappa^2} \ + \delta \vst, \qquad
	V_{\rm I}  = \frac{3 \Delta k_{21}}{\kappa^2} + \delta V_{\rm I} \ , \qquad 
	V_{\rm 1/2} = -\frac{6 k_2}{\kappa^2} + \delta V_{\rm 1/2} \ .
	\label{tensions_shift:eq} 
\end{equation}
Their resulting contributions to the action are
\begin{align}
	\nonumber
  S \supset -\int d^4 x \bigg[ \int_{0}^{\phi_{\rm I}} d\phi \Big(   \sqrt{|g_{*}|}  \delta(\phi) \delta \vst 
&+ \sqrt{|g_{\rm I}|}  \delta(\phi-\phi_{\rm I}) \delta V_{\rm I}  \Big) \\
&+  \int_{\phi_{\rm I}'}^{\phi_{\rm IR}'} d\phi'  \sqrt{|g_{\rm 1/2}|}  \delta(\phi'-\phi_{\rm IR}') \delta V_{\rm 1/2} \bigg]
	\label{action_delta:eq} \ .
\end{align}
Here, $g_{\rm *, I, 1/2}$ are the induced metrics on the UV, intermediate and IR branes, respectively.

We now perform the Kaluza-Klein (KK) decomposition and integrate out the extra dimension
to obtain the 4D effective action for the zero-mode gauge fields. 
From the subregion 1, we find
	\begin{align}
		\nonumber 
		S_{\rm eff} \supset S_{\rm YM, eff}^{(1)} 
										&= \int d^4x 
							\bigg[ - \frac{1}{4 g_{5, 1}^2 k_1} \ln \frac{k_1}{\mu_1} F_{\mu \nu, 1}^{(0)} F^{\mu \nu, 1 (0)} \bigg]  \ ,
										\label{SYM_eqn}
	\end{align}
	where $F_{\mu \nu, 1}^{(0)}$ denotes the field strength of the zero-mode of the $SU(N_{H, 1})$ gauge field. 	
	Taking account of the 1-loop RG flow from the scale $k_1$ to a scale $Q \lesssim \mu_1$,
	the 4D gauge coupling of the zero-mode gauge field for the subregion 1 becomes
		\begin{equation}
		\begin{split}
			\frac{1}{g_{4, 1}^2 (Q, \mu_1)} = \frac{1}{g_{5, 1}^2 k_1} \ln \frac{k_1}{\mu_1}  - \frac{b_{\rm YM, 1}}{8 \pi^2} \ln \frac{k_1}{Q}  \quad ; \quad {\rm for \ Q \lesssim \mu_1} \ .
		\end{split}
		\end{equation}
	Here, $b_{\rm YM, 1} = \frac{11}{3} N_{H, 1}$ for the pure Yang-Mills theory. 
	In the dual 4D picture, the first term in the right hand side can be understood as the contribution from
	the CFT degrees of freedom, which vanishes below the scale $\mu_1$.
	Thus, one can define
		\begin{equation}
			b_{\rm CFT, 1} \equiv -\frac{8 \pi^2}{k_1 g_{5, 1}^2} \quad ; \quad n_1 \equiv -\frac{b_{\rm CFT, 1}}{b_{\rm YM, 1}}  \ .
		\end{equation}
		A natural expectation from the CFT is $b_{\rm CFT, 1} \simeq -\xi_1 N_1$ with a positive constant $\xi_1$.
	The confinement scale is given by the scale where the 4D gauge coupling blows up,
		\begin{align}
			\Lambda_{H, 1} (\mu_1) = k_1  \bigg( \frac{\mu_1}{k_1}\bigg)^{n_1} 
					 \equiv \Lambda_{H, 1}^{(0)} \bigg( \frac{\mu_1}{\mu_{\rm 1, min}} \bigg)^{n_1} 
					 \ , \label{lambda1_conf_scale_mug}
		\end{align}
	where $ \Lambda_{H, 1}^{(0)}$ denotes the confinement scale for $\mu_{\rm 1} = \mu_{\rm 1, min}$.
	This equation is valid only when the confinement scale $\Lambda_{H, 1} (\mu_1) $ is somewhat smaller than
	the radion field value $\mu_1$.
	We parametrize our ignorance of the threshold between the confinement and deconfinement phases
	with $\gamma_{c, 1}$:
	\begin{equation}
			\Lambda_{H, 1} (\mu_{c, 1}) = \gamma_{c, 1} \mu_{c, 1} \ .
		\end{equation} 
	Here, the scale $\mu_{c, 1}$ is defined such that Eq.~\eqref{lambda1_conf_scale_mug} is valid for $\mu_1 \ge \mu_{c, 1}$.
	As the 4D effective description breaks down for the confinement scale larger than the mass of the lightest KK mode, one expects $\gamma_{c, 1} \simeq \pi$.
	For $\mu_1 < \mu_{c, 1}$, on the other hand, we expect the confinement scale to become independent of the radion field value,
		\begin{equation}
			\Lambda_{H, 1} (\mu_1) = \gamma_{c, 1} \mu_{c, 1} \quad ; \quad {\rm for \ \mu_1 < \mu_{c, 1}} \ .
			\label{lambda1_conf_scale_mul}
		\end{equation}
	Equating Eq.~\eqref{lambda1_conf_scale_mug} with Eq.~\eqref{lambda1_conf_scale_mul} at the scale $\mu_{c, 1}$, 
	we find
			\begin{equation}
			\mu_{c, 1} = \mu_{\rm 1, min} \bigg(  \frac{\Lambda_{H, 1}^{(0)}}{\gamma_{c, 1} \mu_{\rm 1, min}} \bigg)^{1/(1-n_1)} \ .
			\label{muc1_eqn}
		\end{equation}
	The dark gluon condensates contribute to the heavier radion potential,
		\begin{equation}
			V_{H, 1} (\mu_1) = \frac{1}{4} \langle T_{1, \mu}^\mu \rangle \simeq -\frac{b_{\rm YM, 1}}{8} \Lambda_{H, 1}^4(\mu_1) \ ,
		\end{equation}
	where the non-vanishing trace of the stress-energy tensor $T_{1, \mu}^{\mu}$ appears as a result of the conformal anomaly. Therefore, the total effective potential for the heavier radion is given by 
		\begin{equation}
		V_{\rm r, eff, 1} (\mu_1)= 	\begin{cases}
													\delta \vst + \frac{\lambda_1}{4} \mu_1^4 - \frac{b_{\rm YM, 1}}{8} (\Lambda^{(0)}_{H, 1})^4 \bigg(\frac{\mu_1}{\mu_{\rm 1, min}}\bigg)^{4 n_1} \quad ; \quad {\rm for} \ \mu_1 >  \mu_{c, 1} \\
													\delta \vst + \frac{\lambda_1}{4} \mu_1^4 - \frac{b_{\rm YM, 1}}{8} \gamma_{c, 1}^4 \mu_{c, 1}^4 \quad ; \quad {\rm for} \ \mu_1 < \mu_{c, 1} \ .
										\end{cases}
										\label{heavier radion potential}
		\end{equation}
	Here, $\lambda_1 \equiv 4 \delta V_{\rm I}/k_1^4$, and this quartic coupling of $\mu_1$ can be balanced
	against the term generated from the condensation, leading to a non-trivial minimum for the heavier radion.
	For $n_1 < 1$, the radion potential has a global minimum,
		\begin{align}
			\nonumber
			\mu_{\rm 1, min} &={} \Big( \frac{n_1 b_{\rm YM, 1}}{2 \lambda_1}\Big)^{1/4} \Lambda_{H, 1}^{(0)} \ , 
			\\[1ex]
			V_{\rm r, eff, 1}(\mu_{\rm 1, min}) &={}  -\frac{\lambda_1}{4}\Big(\frac{1}{n_1}-1 \Big) \mu_{\rm 1, min}^4 \ .
			\label{Vmin_dark_qcd:eq}
		\end{align}
	The constant $ \delta \vst$ in Eq.~\eqref{heavier radion potential} can be chosen to reproduce
	the observed vanishingly-small cosmological constant.

	As the heavier radion $\mu_1$ has been stabilized at $\mu_1 = \mu_{\rm 1, min}$, we look at the evolution of the theory from there on. At the second stage of the evolution, the $SU(N_{H, 2})$ gauge field in the subregion 2 confines,
	and similarly, we obtain the 4D effective action for the zero-mode gauge field,
	\begin{align}
	\nonumber 
	S_{\rm eff} \supset S_{\rm YM, eff}^{(2)} &= \int d^4 x     \int_{\phi_{\rm I}'}^{\phi_{\rm IR}'} T_2 (x) d\phi'  \bigg( -\frac{1}{4 g_{5, 2}^2} F_{\mu \nu, 2}^{(0)} F^{\mu \nu, 2 (0)}\bigg) \ .
	\label{SYM_eqn2}
\end{align}
 From the definition in Eq. (\ref{radion_defns}), we have the following relation:
		\begin{equation}
			T_2(x) (\phi_{\rm IR}'-\phi_{\rm I}') = \frac{1}{k_2} \ln \bigg( \frac{\mu_1(x)}{\mu_2(x)} \frac{k_2}{k_1}\bigg) \ .
		\end{equation}
	Hence, the 4D gauge coupling of the zero-mode of the $SU(N_{H, 2})$ gauge field at a scale $Q \lesssim \mu_2$ becomes
		\begin{equation}
			\frac{1}{g_{4, 2}^2 (Q, \mu_2)} = \frac{1}{k_2 g_{5, 2}^2} \ln \frac{k_2}{\mu_2} + \frac{1}{k_2 g_{5, 2}^2} \ln \frac{\mu_{\rm 1, min}}{k_1}  - \frac{b_{\rm YM, 2}}{8 \pi^2} \ln \frac{\mu_{\rm 1, min}}{Q} \quad ; \quad {\rm for } \ Q \lesssim \mu_2 \ ,
		\end{equation}
		with $b_{\rm YM, 2}  = (11/3) N_{H, 2}$.
		Note that the heavier radion $\mu_1$ has been already stabilized, and $g_{4, 2}$ does not depend on $\mu_1$ dynamically but just on $\mu_{\rm1,  min}$. 
		 The confinement scale for the $SU(N_{H, 2})$ is then given by
		\begin{equation}
			\Lambda_{H, 2} (\mu_2) = \Lambda_{H, 2}^{(0)} \bigg( \frac{\mu_2}{\mu_{\rm 2, min}}\bigg)^{n_2} \ ,
		\end{equation}
	where
		\begin{equation}
			b_{\rm CFT, 2} \equiv -\frac{8 \pi^2}{k_2 g_{5, 2}^2} \quad ; \quad
			n_2 \equiv - \frac{b_{\rm CFT, 2}}{b_{\rm YM, 2}} \ .
		\end{equation}
		As in the case of the $SU(N_{H, 1})$, the dark gluon condensates lead to the lighter radion potential.
		In addition, the deviation of the IR brane tension $\delta V_{\rm 1/2}$ generates an effective potential of the form -$\frac{\lambda_2}{4} \mu_2^4$ with $\lambda_2 = 4 \delta V_{\rm 1/2}/k_2^4$ for the lighter radion. 
	Therefore, the similar discussion for the stabilization of $\mu_2$ follows as long as $n_2 <1$.

	\subsection{Phase transitions}
	
	As discussed in section~\ref{phase_transition_section},
	our three 3-brane setup at finite temperature can have three regimes:
	(i) the AdS-S spacetime with an event horizon, (ii) the AdS-S spacetime with the intermediate 3-brane and (iii) the time-compactified RS spacetime with the intermediate and IR branes.
	Then, we expect the phase transition from the regime (i) to (ii) and the transition from the regime (ii) to (iii),
	as described in figure~\ref{fig:flow_diagram},
	and need to investigate whether those phase transitions are correctly completed
	in the current model of radion stabilization with two dark gauge fields. 
	
	Let us first consider the phase transition from the regime (i) to (ii).
For $T\ll T_{c,1}$, the dominant contribution to the $O(4)$-symmetric bounce action comes from the gradient energy
and is reliably estimated by the thick-wall approximation~\cite{Nardini:2007me},
	 	\begin{equation}
	 		S_4^{(1)} \simeq \frac{9 N_1^4}{8 \pi^2} \frac{(\mu_{t,1} + T)^4}{V_{\rm r, eff, 1}(\mu_{\rm 1, min}) \Big(\frac{T}{T_{c,1}}\Big)^4 - V_{\rm r, eff, 1}(\mu_{t,1})} \ ,
	 	\end{equation}
	 where the critical temperature $T_{c,1}$ in Eq.~\eqref{critical_temp:eq} is given by
	 	\begin{equation}
	 		T_{c,1} = \bigg(8  \frac{|V_{\rm r, eff, 1}(\mu_{\rm 1, min})|}{\pi^2 N_1^2} \bigg)^{1/4} \ .
	 	\end{equation}
	The tunneling point $\mu_{t,1}$ is found by minimizing the action,
	$\frac{\partial S_{4}^{(1)}}{\partial \mu_{t, 1}} = 0$,
	and for a low $T$, it is obtained as
		\begin{equation}
			\mu_{t, 1} = \mu_{\rm 1, min} \bigg(\frac{1}{1-n_1}\bigg)^{\frac{1}{4 n_1}} \bigg(\frac{2 \lambda_1}{\gamma_{c, 1}^4 n_1 b_{\rm YM, 1}}\bigg)^{\frac{1}{4(1-n_1)}} \ .
		\end{equation}
	If the phase transition is correctly completed, the nucleation temperature, $T_{n, 1}$,
	is given by the solution of the equation,
	\begin{equation}
	\Gamma_{n} = \mu_{t, 1}^4 e^{-S_4^{(1)}} \simeq H^4 \ ,
	\end{equation}
	with the Hubble rate $H \simeq \mu_{\rm 1, min}^2/M_{\rm pl}$.
	
\begin{figure}[t!]
		\begin{center}
		\hspace{-0.3cm}
			\begin{subfigure}{0.48\textwidth}
				\centering
				\includegraphics[width=\textwidth]{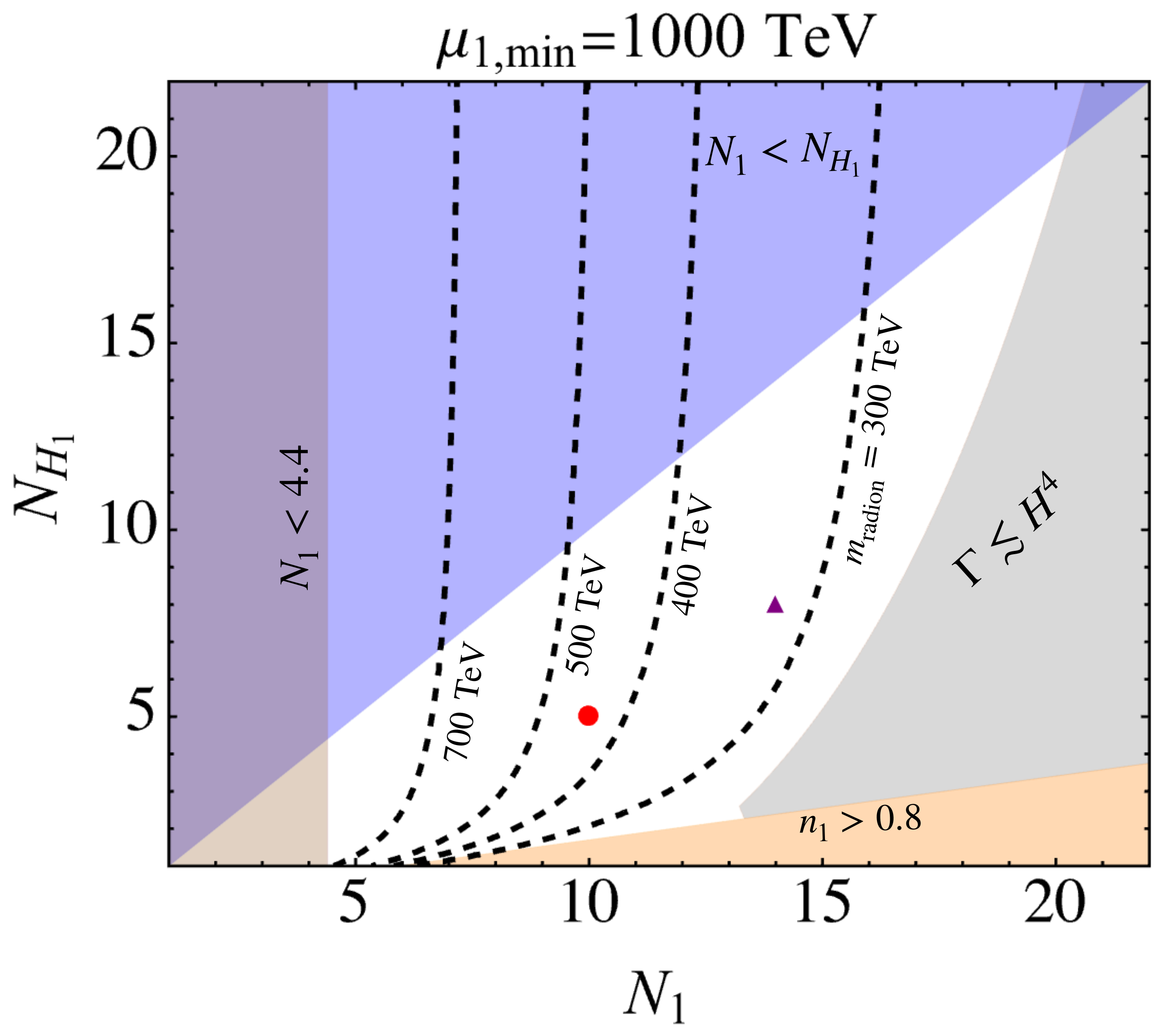}
				\subcaption{}
				\label{}
			\end{subfigure}
			\hfill
			\begin{subfigure}{0.48\textwidth}
				\centering
				\includegraphics[width=\textwidth]{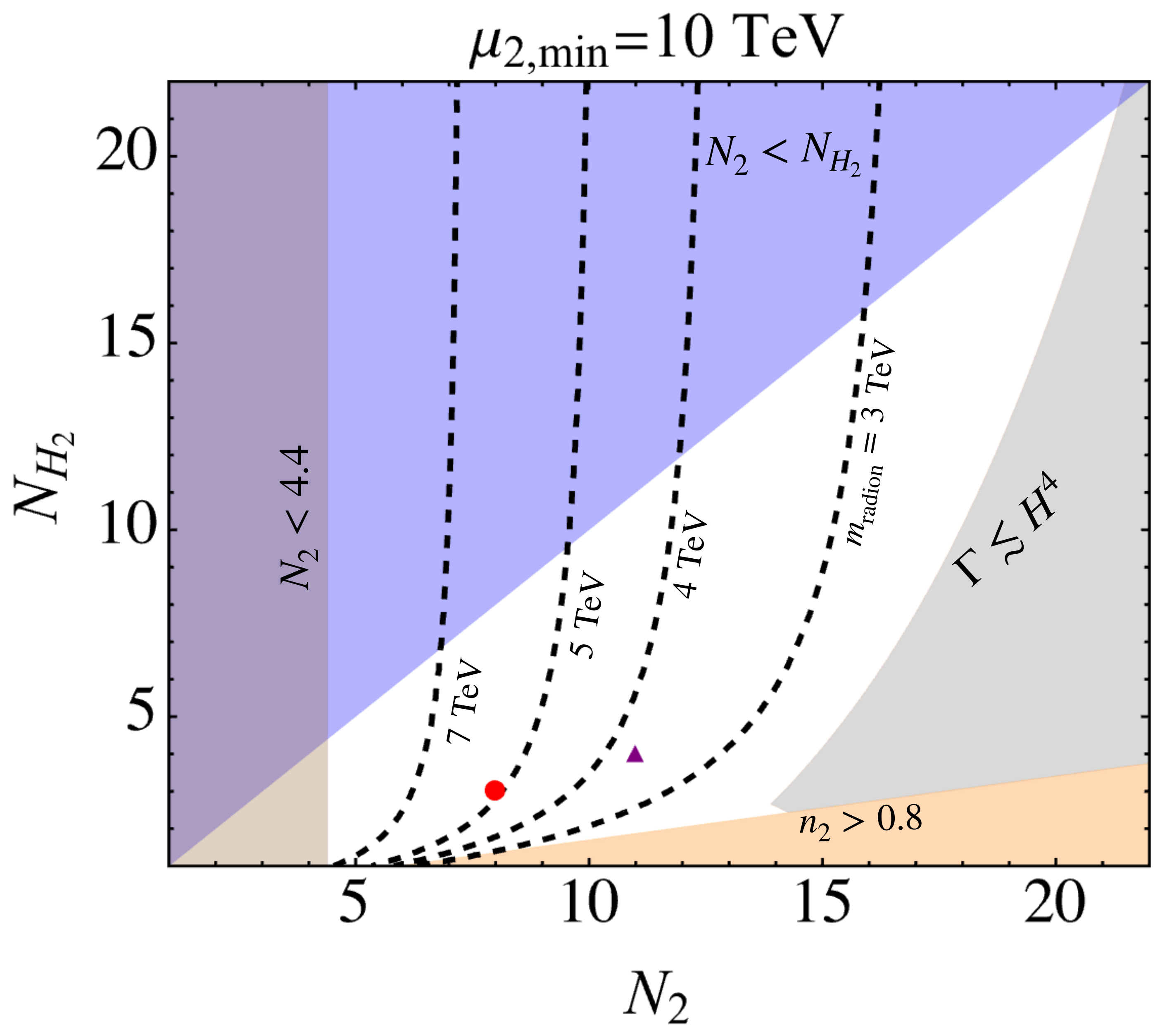}
				\subcaption{}
				\label{}
			\end{subfigure}
		\end{center}
		\vspace{-0.3cm}
		\caption{The allowed (white) parameter regions that the phase transition from the regime (i) to (ii) (left)
		and the transition from (ii) to (iii) (right) get completed in the model with dark gauge fields. In each panel, the gray shaded region shows
		the excluded region that the bubble nucleation action is too large to compete with the Hubble rate.
		A cautionary remark is that the boundary of the gray shaded region is only an analytical estimate of the zero-temperature action.
		The blue shaded regions correspond to $N_{H_i} > N_i$, where the finite temperature effects of dark gauge fields become important (this is an approximate estimation, and can be modified by ${\cal O}(1)$ correction coefficient). The orange regions, corresponding to $n_i > 0.8$, have $\mu_{c, i}$ smaller than the QCD scale, and its effect has to be included. The left most violet exclusion regions correspond to the regime where quantum gravity effects become important. The red circle and violet triangle in each panel denote benchmark points for the discussion of the gravitational wave production (see figure~\ref{gw_fig}).
		In the left and right panels, the black dashed contours depict the heavier and lighter radion masses, respectively.
		The remaining parameters are chosen as $\lambda_{1, 2} = 1$, $\xi_{1,2} = 0.5$, and $\gamma_{c_{1, 2}} = \pi$.} 
		\label{paraspace_dark_qcd}
	\end{figure}

	The left panel of figure~\ref{paraspace_dark_qcd} shows the parameter region
	that the bounce action is small enough to compete with the Hubble rate and hence
	the phase transition from the regime (i) to (ii) is safely completed.
	We can see that there exists a large viable region in the current model, unlike the case of the GW mechanism.
	The underlying reason for this result can be understood as follows. The radion potential generated by the GW mechanism is shallow as it is suppressed by $\epsilon$ in Eqs.~\eqref{vmin_GW_delta:eq}, \eqref{Vmin_GW_epsilon:eq}, and $\epsilon \ll 1$ in order to produce a hierarchically smaller mass scale than the Planck scale.
	Furthermore, one cannot make the VEVs of the GW scalar on the branes arbitrarily large without violating the backreaction condition in Eq.~(\ref{backreaction_eqn}). In contrast, from Eq.~(\ref{Vmin_dark_qcd:eq}), it is evident that the radion potential generated via the dark gauge field confinement can be much deeper, and hence can allow for a smaller bubble nucleation action. In terms of the dual CFT picture, the breaking of the scale invariance by a small parameter $\epsilon$ results into a much shallower dilaton potential for the GW-type stabilization, whereas the dark gauge field confinement breaks the scale invariance strongly at low energies. The backreaction problem is circumvented as well, thanks to the asymptotically-free nature of the dark gauge field.

	%
	%

	After the first phase transition is completed, the intermediate brane appears and the Universe gets reheated.
	Assuming that the vacuum energy in the false vacuum is transformed into the radiation in the true vacuum
	and using the definition of the critical temperature in Eq.~(\ref{critical_temp:eq}),
	the reheating temperature is estimated as
	\begin{equation}
	\begin{split}
			T_{\rm RH, 1} \simeq \bigg(\frac{45}{4}\bigg)^{1/4} \frac{\sqrt{N_1}}{g_\ast^{1/4} (T_{\rm RH, 1})} T_{c, 1} \ ,
			\end{split}
	\end{equation}
	where $g_\ast (T_{\rm RH, 1})$ denotes the number of degrees of freedom at $T_{\rm RH, 1}$.
	As the reheating temperature is above the energy scale of the IR brane,
	we enter into the phase where the IR brane is replaced by a horizon.
	The minimum of the free energy in this phase increases as the Universe is cooled, and eventually, becomes comparable to the potential minimum for the lighter radion, which defines as before the critical temperature $T_{c,2}$ for the second phase transition from the regime (ii) to (iii).
	The similar argument as in the case of the phase transition from the regime (i) to (ii) can be applied to
	the second transition.
	In the right panel of figure~\ref{paraspace_dark_qcd}, we show the parameter region
	that the bounce action is small enough to compete with the Hubble rate and the phase transition from the regime (ii) to (iii) is completed.

Before closing this section, let us mention about the possibility for the production of dark glueballs,
which are formed below the confinement scales of the dark gauge fields.
However, the produced glueballs are unstable and immediately decay into lighter particles in the thermal bath.
In more detail, let us first consider dark glueballs from the subregion 1. They, for example, couple to
(zero-mode) dark gluons in the subregion 2 by
\begin{align}
    \mathcal{L}_{\rm eff} \supset \frac{\Lambda_{H,1}^{(0)\,3}}{4\pi M_{\rm I}^4}\Phi_1 F_{\mu \nu, 2}^{(0)} F^{\mu \nu, 2 (0)}\ ,
\end{align}
where $M_{\rm I}\sim k_1 e^{-k_1 y_{\rm I}}$, $\Phi_1$ denotes the dark glueballs from the subregion 1
and the coupling constant has been omitted.
With $M_{\rm I}\gtrsim \Lambda_{H,1}^{(0)}$, the glueballs $\Phi_1$ decay into lighter particles immediately
and do not remain in the Universe.
Similarly, dark glueballs from the subregion 2 couple to the SM particles, $e.g.$
through interactions between the dark glueballs and the photons,
and decay quickly.

	\section{Gravitational wave generation}
	
	The first order phase transitions in our multi-brane model produce stochastic gravitational wave backgrounds.
	We now assume the radion stabilization mechanism with dark gauge fields,
	discussed in section~\ref{dark_qcd_stabilization},  and estimate their energy densities.
	Following ref.~\cite{Lisa2}, a strong first-order phase transition can be described in terms of two quantities: the latent heat released $\alpha$, which is defined as the ratio of the vacuum energy density released $\rho_{\rm vacuum}$
	to the radiation density $\rho_{\rm rad} $ at the bubble nucleation temperature $T_n$, namely
	\begin{equation}
		\alpha \equiv \frac{\rho_{\rm vacuum}}{\rho_{\rm rad} (T_n)} \simeq \frac{| \Delta F |}{\rho_{\rm rad}(T_n)} \gg 1 \ ,
	\end{equation}
where $|\Delta F |$ is the free energy difference between the false and true vacua, and the inverse duration of the phase transition,
	\begin{align}
		\nonumber
		\frac{\beta}{H(T_{\rm RH})} & 
		\equiv -\frac{dS_4}{dt}  \Big|_{t = t_\ast} \simeq \frac{\dot{\Gamma}}{\Gamma}  
		 \simeq  \frac{H(T_n)}{H({T_{\rm RH}})}T_n \bigg(\frac{d S_4}{d T}\bigg)_{T_n}   \ .
	\end{align}
	Here,  $T_{\rm RH}$ is the reheating temperature, and $t_\ast$ denotes the time when gravitational waves are produced.
	We assume that $t_\ast$ is approximated by the time corresponding to the nucleation temperature.
	One feature of the current radion stabilization mechanism is that the nucleation temperature is not much smaller than the critical temperature. This means that a supercooling phase between the critical and nucleation temperatures is short compared to the corresponding one from the Goldberger-Wise stabilized model,
	so that gravitational waves generated from the first transition from the regime (i) to (ii) are not completely diluted away during the second transition from the regime (ii) to (iii) \footnote{see also~\cite{Ellis:2018mja} for a discussion on supercooled phase transition in warped models and its compatibility with successful bubble percolation.}.
	Since the second phase transition also produces gravitational waves, the total gravitational wave density is estimated as
		\begin{equation}
			\Omega_{\rm GW} h^2 \simeq  \bigg(\frac{g_{*,s}(T_{n, 2})}{g_{*,s}(T_{\rm RH, 2})}\bigg)^{4/3} \bigg(\frac{T_{n,2}}{T_{\rm RH, 2}}\bigg)^4 \Omega_{\rm GW, 1} h^2 + \Omega_{\rm GW, 2} h^2 \ ,
		\end{equation}
	where $g_{*,s}$ denotes the effective number of degrees of freedom in entropy, the two terms in the right hand side come from the two consecutive  phase transitions,
	and we have assumed instantaneous reheating.
	Clearly, if we had $T_{n,2} \ll T_{\rm RH,2}$, the gravitational wave density generated from the first transition would be significantly suppressed.
\begin{table}[t]
	\resizebox{\linewidth}{!}{%
		\centering
		    \begin{tabular}{||c|c|c|c|c|c|c|c|c|c|c|c|c|c||}
			\hline
			{\rm fig.} & {\rm curve} & {$\mu_{\rm 1, min}$} & {$\mu_{\rm 2, min}$} & $T_{c,1}$ & $T_{n,1}$ & $T_{\rm RH,1}$ & $T_{c,2}$ & $T_{n,2}$ & $T_{\rm RH,2}$ & $\alpha_{1}$ & $\left(\beta/H\right)_1$ & $\alpha_2$ & $\left(\beta/H\right)_2$  \\ 
			\hline
			{\rm 4} & {\rm blue solid} & 1000 &  10 &  271 & 128 & 497 & 3 & 1.3 & 5 & 15 & 19 & 17 & 50\\ 
			{\rm 4} & {\rm black dashed} & 1000 &  10 & 240 & 73 & 519 & 2 & 0.8 & 4 & 82 & 3 & 76 & 8\\ 
			{\rm 5} & {\rm Magenta solid} & 50 &  3 & 4 & 1 & 9 & 1 & 0.7 & 2 & 307 & 2 & 2 & 71 \\
			{\rm 5} & {\rm black dotdashed} & 50 &  3 & 4 & 0.8 & 8 & 1 & 0.7 & 2 & 568 & 2 & 2 & 152 \\
			\hline
	\end{tabular}}
	\caption{Relevant temperature scales (in TeV) and gravitational wave parameters for the chosen benchmark points in figs.~\ref{gw_fig},~\ref{gw_fig_llisa}. The subscript $i=1, 2$ refers to the corresponding parameter for the $i^{\rm th}$ phase transition.}
	\label{tab:alpha_beta}
\end{table}
	The gravitational wave density produced by each of the two phase transitions can be decomposed into contributions
	from the bubble collision, sound wave and turbulence of the thermal plasma, namely
		\begin{equation}
			\Omega_{{\rm GW}, i} h^2 = \Omega_{{\rm col}, i} h^2 + \Omega_{{\rm sw}, i} h^2 + \Omega_{{\rm turb}, i} h^2 \quad ; \quad i = 1,2 \ .
		\end{equation}
	The background of the strongly interacting $SU(N_{H, i})$ gauge field provides the thermal plasma that results in a terminal velocity for the expanding bubble wall~\cite{Fujikura:2019oyi, Lisa2}. Therefore, the dominant contribution comes from the bulk motion of the fluid, in the form of the sound wave and turbulence.
	We use the numerical approximation and modeling presented in ref.~\cite{Lisa2}.
	The contribution of the sound wave is then given by
	
	\label{gw_section}
	\begin{figure}[t!]
		\includegraphics[width=0.75\textwidth]{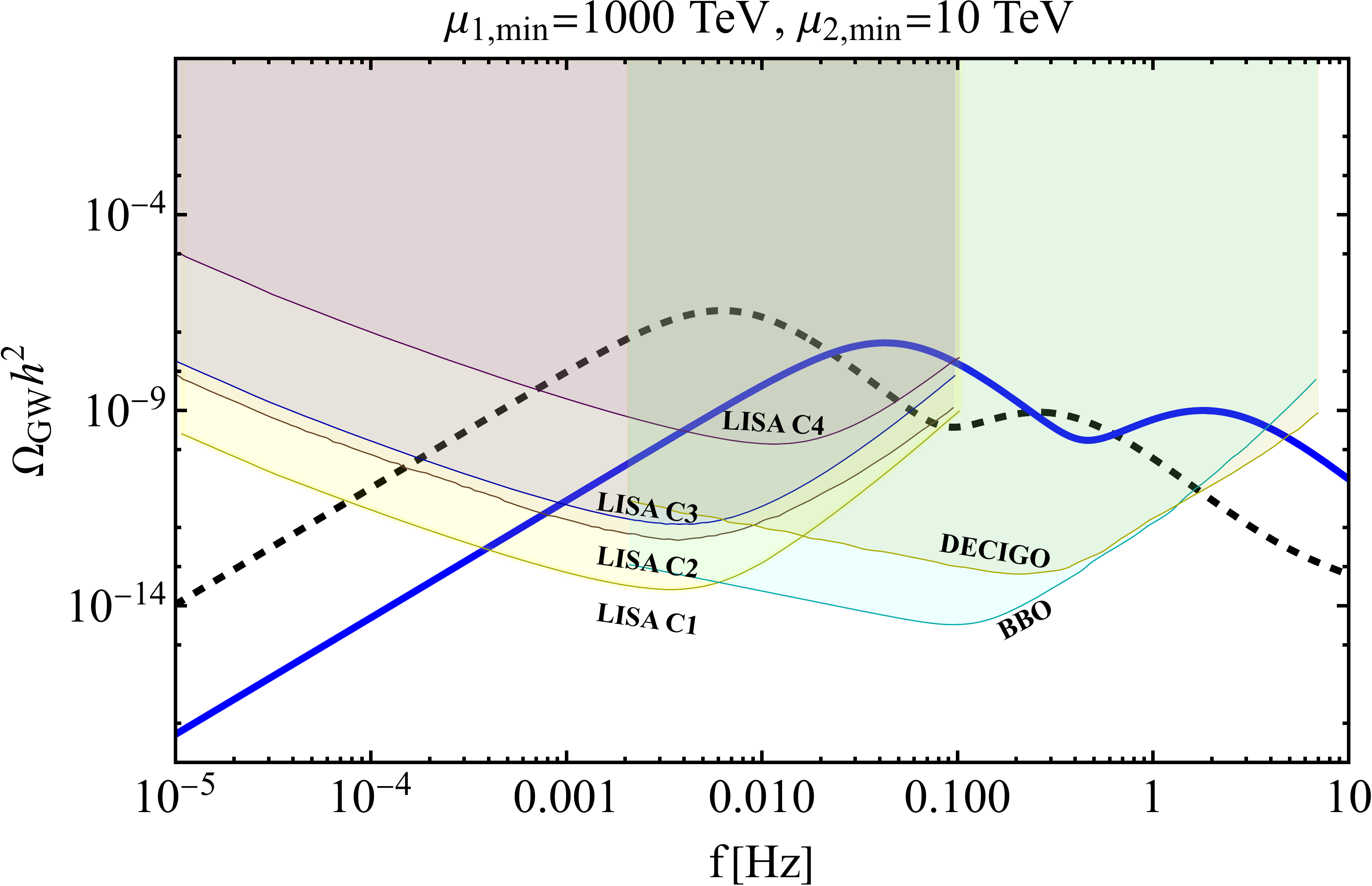}
		\vspace{0.3cm}
		\caption{Relic abundance of gravitational waves from two successive first-order phase transitions
		in the multi-brane model with radions stabilized by dark gauge fields.
		The blue solid line corresponds to the case with $N_1=10$, $N_{H_1} = 5$, $N_2=8$ and $N_{H_2} = 3$
		(red circles in figure~\ref{paraspace_dark_qcd}), whereas the black dashed line corresponds to the case with $N_1=14$, $N_{H_1} = 8$, $N_2=11$ and $N_{H_2} = 4$ (violet triangles).
		The intermediate and IR brane mass scales are 1000 and 10 TeV, respectively.
		The other parameters are chosen as $\lambda_{1, 2} = 1$, $\xi_{1,2} = 0.5$ and $\gamma_{c_{1, 2}} = \pi$.
		}
		\label{gw_fig}
	\end{figure}
	%
	\begin{figure}[t!]
		\includegraphics[width=0.75\textwidth]{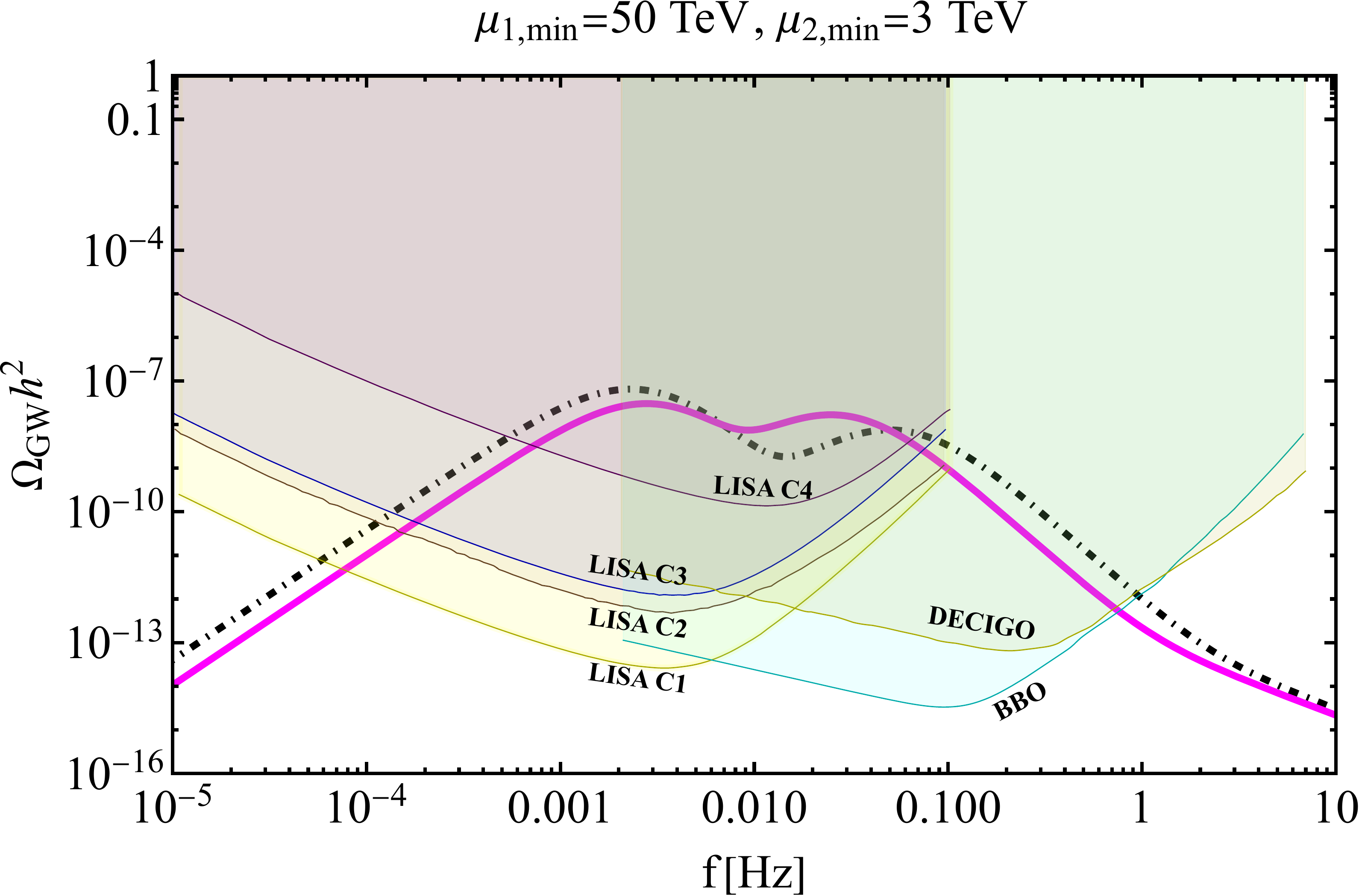}
		\vspace{0.3cm}
		\caption{The similar plot as that of figure~\ref{gw_fig}, with different choices of the parameters:
		$N_1=13$, $N_{H_1} = 4$, $N_2=10$ and $N_{H_2} = 8$ for the magenta solid curve, and $N_1=12$, $N_{H_1} = 3$, $N_2=8$ and $N_{H_2} = 5$ for the black dot-dashed curve.  The intermediate and IR brane mass scales are 50 and 3 TeV, respectively. The other parameters are chosen as $\lambda_{1, 2} = 1$, $\xi_{1,2} = 0.4$ and $\gamma_{c_{1, 2}} = \pi$.}
		\label{gw_fig_llisa}
	\end{figure}
	
	\begin{align}
		\nonumber
		\Omega_{\mathrm{sw}, i }(f) h^2 & \simeq 2.65 \times 10^{-6} \times\left(\frac{H\left(T_{\mathrm{RH}, i}\right)}{\beta_i}\right)\left(\frac{\kappa_{\mathrm{sw}} \alpha_i}{1+\alpha_i}\right)^2\left(\frac{100}{g_*}\right)^{\frac{1}{3}} v_w\left(\frac{f}{f_{\mathrm{sw}, i}}\right)^3\left(\frac{7}{4+3\left(\frac{f}{f_{\mathrm{sw}, i}}\right)^2}\right)^{\frac{7}{2}} , \\[1ex]
			f_{\mathrm{sw}, i} & \simeq 1.9 \times 10^{-4} \mathrm{~Hz} \times \frac{1}{v_w}\left(\frac{\beta_i}{H\left(T_{\mathrm{RH}, i}\right)}\right)\left(\frac{T_{\mathrm{RH}, i}}{1 \, \mathrm{TeV}}\right)\left(\frac{g_*}{100}\right)^{\frac{1}{6}} ,
			\label{gw_sw_eqn}
	\end{align} 
where $i = 1, 2$. For our purposes, we set the efficiency factor $\kappa_{\mathrm{sw}} \simeq 1$
and the bubble wall velocity $v_w \simeq 1$. Note that as $\alpha_i \gg 1$, its dependence drops out in Eq.~(\ref{gw_sw_eqn}). $f_{\mathrm{sw}}$ denotes the peak frequency, redshifted accordingly.
The contribution of the turbulence is
\begin{align}
\nonumber
\Omega_{\mathrm{turb}, i}(f) h^2 & \simeq 3.35 \times 10^{-4} \times\left(\frac{H\left(T_{\mathrm{RH}, i}\right)}{\beta_i}\right)\left(\frac{\kappa_{\mathrm{turb}} \alpha_i}{1+\alpha_i}\right)^{\frac{3}{2}}\left(\frac{100}{g_*}\right)^{\frac{1}{3}} v_w \frac{\left(\frac{f}{f_{\text {turb}, i}}\right)^3}{\left(1+\frac{f}{f_{\text {turb}, i}}\right)^{\frac{11}{3}}\left(1+\frac{8 \pi f}{h^{*}_i}\right)}  \ ,
\\[1ex]
\nonumber
h^{*}_{i} & \simeq 1.65 \times 10^{-4} \mathrm{~Hz} \times\left(\frac{T_{\mathrm{RH}, i}}{1 \, \mathrm{TeV}}\right)\left(\frac{g_*}{100}\right)^{\frac{1}{6}} \ , 
\\[1ex]
f_{\text {turb},i} & \simeq 2.7 \times 10^{-4} \mathrm{~Hz} \times \frac{1}{v_w}\left(\frac{\beta}{H\left(T_{\mathrm{RH}, i}\right)}\right)\left(\frac{T_{\mathrm{RH}, i}}{1 \, \mathrm{TeV}}\right)\left(\frac{g_*}{100}\right)^{\frac{1}{6}} \ .
\label{gw_turb_eqn}
\end{align}
We take $\kappa_{\mathrm{turb}} \simeq 0.05 \kappa_{\rm sw}$, conservatively. For more details on these analytical approximations, we point the reader to ref.~\cite{Lisa2} and references therein. 

Figure~\ref{gw_fig} shows the gravitational wave spectrum with unique two peaks predicted in our model. The peak towards the higher characteristic frequency comes from the first phase transition from the regime (i) to (ii). In table~\ref{tab:alpha_beta} we show the relevant temperature scales and gravitational wave parameters for the chosen benchmark points. Note that this peak is damped down due to the supercooling during the second phase transition, while this is not washed away hopelessly
beyond the sensitivity limit of future gravitational wave probes such as LISA, DECIGO, and BBO.
If the intermediate and IR brane mass scales lie at relatively lower scales, then
the associated reheating temperatures become smaller, and with appropriate choices of the parameters in the viable region,
the two peaks can lie in the sensitivity range covered by LISA, as shown in figure~\ref{gw_fig_llisa}.

\section{Discussions}\label{conclusion}

Warped extra dimensional models with multiple 3-branes can naturally accommodate hierarchically different mass scales that often appear in physics beyond the SM. We have contemplated the cosmological effects of such multi-brane warped models. In particular, it was confirmed that the usual FLRW Universe is recovered below the IR mass scale, and the Hubble expansion parameter is determined by the sum of 
brane-localized energy densities. We have also analyzed the first-order confinement-deconfinement phase transitions in the three 3-brane models and pointed out that the problem of the completion of the phase transitions gets worse, in comparison with the case of the two 3-brane model, with the usual GW stabilization mechanism. In hindsight, this is expected, as the bubble nucleation rate has to compete with the even larger Hubble rate corresponding to the intermediate mass scale. To circumvent the problem, we first developed an extension of dark gauge field stabilization of multiple branes. Here we utilized two dark gauge fields, which reside in the bulk subregions 1, 2, respectively. The confinement of these gauge fields provides the strong breaking of the scale invariance, and generates a potential for the radions deeper than that of the GW mechanism, which allows for the phase transitions to get completed. We  have then analyzed the spectrum of gravitational waves generated by the first-order phase transitions. With the radion stabilization with two dark gauge fields, the supercooling epoch can be small enough to allow for the possibility of a two peak gravitational wave frequency spectrum whose amplitude is within the projected sensitivity limits of future space-based gravitational wave observers such as LISA, DECIGO and BBO for generic choices of the parameters. 

In the present paper, we have assumed that the two phase transitions are decoupled from each other. This assumption is valid as long as $N_1 > N_2$, and $\mu_{\rm 1, min} \gg \mu_{\rm 2, min}$. In the dual 4D picture, it can be understood by the fact that the mixing between the two corresponding dilatons are proportional to $\left(\mu_{\rm 2, min}/\mu_{\rm 1, min}\right)^3$. It would be interesting to consider a scenario where the two radions are not completely decoupled from each other, and therefore, it is conceivable that the dynamics of the two phase transitions would be phenomenologically richer in terms of gravitational wave signatures. We leave it for a future study. 

When the system is in the regime (ii), we considered the AdS-S spacetime with the Euclidean time
compactified on a circle, and did not keep track of its time evolution as we have 
discussed in section~\ref{rs3_section} where the system is at a temperature below the scale of the IR brane.
However, it may be important to understand the time evolution of the background spacetime in the regime (ii),
if the system accommodates a topological object, such as a cosmic string,
formed by some spontaneous symmetry breaking at the intermediate brane or in the subregion 1.
The behavior of such a topological object depends on how the background spacetime evolves in time. 
Furthermore, it is interesting to note that a network of cosmic strings acts as a durable source of gravitational waves from the time of their production so that the resulting gravitational wave background stretches across a wide range of frequencies,
which encodes the evolution of  the background spacetime.
The second phase transition from the regime (ii) to (iii) must also affect the evolution of cosmic strings
and their production of gravitational waves.
Since the confinement-deconfinement phase transitions themselves generate gravitational waves,
the final spectrum is given by the mixture of those from cosmic strings and the phase transitions.
We may discover the early Universe based on our multi-brane model
by looking at the frequency spectrum of gravitational waves!

\section*{Acknowledgements}

We would like to thank Kohei Fujikura and Ryo Namba for discussions.
YN is supported by Natural Science Foundation of China under grant No. 12150610465.
This work was supported by JSPS KAKENHI Grant Numbers JP22J00537 (M.S.).
The research activities of SL were supported in part by the National Research 
Foundation of Korea (NRF) grant funded by the Korean government (MEST) 
(No. NRF-2021R1A2C1005615) and the Samsung Science \& Technology Foundation 
under Project Number SSTF-BA2201-06.
%
	
\appendix

\section{Not one but two radions}
\label{two_radions}

As discussed in the main text, there are two radions in the three 3-brane setup. 
In ref.~\cite{Lee:2021wau}, some of the authors of the present paper have argued the existence of $N-1$ radions
in the $N$ 3-brane setup. However, in refs.~\cite{Cai:2021mrw,Cai:2022geu},
the author has claimed that only one radion exists in the same $N$ 3-brane setup, which contradicts our conclusion. 
A relevant difference between ref.~\cite{Lee:2021wau} and refs.~\cite{Cai:2021mrw,Cai:2022geu} is the boundary condition for $E'$ (in the notation of ref.~\cite{Lee:2021wau} 
and $\epsilon'$ in the notation of refs.~\cite{Cai:2021mrw} \cite{Cai:2022geu})
on the intermediate brane.
In refs.~\cite{Cai:2021mrw,Cai:2022geu}, $E' =0$ is eventually taken on the intermediate brane
to identify the radion degrees of freedom,
while
$E’\neq 0$ is generally allowed on the intermediate brane, $i.e.$ if the brane is {\it not} at the fixed points, the jump condition and the symmetry of the system do {\it not} require $E’=0$ on the brane. 
Indeed, in ref.~\cite{Lee:2021wau}, 
$E’\neq 0$ on the intermediate brane is derived from the continuity of $F$ and $E’$ on that brane. 
In refs.~\cite{Cai:2021mrw,Cai:2022geu}, it is unclear  {where the condition of $E’=0$ on the intermediate brane comes from.}
Ref.~\cite{Cai:2021mrw}  simply takes it for granted that $E’=0$ condition is correct, and the author worries about an ambiguity
due to a nonzero $E'$, while ref.~\cite{Lee:2021wau} does not encounter such an ambiguity.
In more detail, around Eq.~(6.4) in ref.~\cite{Lee:2021wau}, the authors get proper radion kinetic terms without any ambiguity for $E’$ on the intermediate brane.
Ref.~\cite{Cai:2022geu} argues that if the radion is not stabilized, $E’ \neq 0$ on the intermediate brane can be taken.
Following the discussion of ref.~\cite{Lee:2021wau}, however, this {implies} that another radion degree of freedom appears in the three 3-brane setup, at least, for the case without radion stabilization.
This statement  {cannot be correct} because the number of radion degrees of freedom should be determined regardless of their stabilization. In ref.~\cite{Lee:2021wau}, the condition of $E’ \neq 0$ is crucial to see multiple radion degrees of freedom,
$i.e.$ if we consider $E’= 0$, we can only see one radon degree of freedom.
Ref.~\cite{Kogan:2001qx} has also discussed that $E'$ on each brane is gauge independent,
and a non-trivial $E'$ is obtained on the intermediate brane.

Another crucial difference between ref.~\cite{Lee:2021wau} and refs.~\cite{Cai:2021mrw,Cai:2022geu}
appears in the radion mass computation. 
In Eq.~(34) of ref.~\cite{Cai:2021mrw}, an incomprehensible boundary condition on the intermediate brane is imposed.
Apparently, this condition contradicts Eq.~(5.9) and Eq.~(5.10) of ref.~\cite{Lee:2021wau}
($f_{1,2}$ in ref.~\cite{Cai:2021mrw} correspond to $d_{1,2}$ in ref.~\cite{Lee:2021wau}).

{Let us consider what would happen if $E'=0$ were taken by hand on the intermediate brane.} Then, we would get the same forms of the bulk equation and the boundary condition in every subregion as those for the two 3-brane case, $i.e.$ the form of the bulk equation is the same as Eq.~(6.3) of ref.~\cite{Csaki:2000zn}
and the form of the boundary condition is the same as Eq.~(6.4) of ref.~\cite{Csaki:2000zn}.
If we use the same forms of the bulk equation and the boundary condition in every subregion
but take $e.g.$ different bulk cosmological constants in different subregions ($i.e.$ $k_1\neq k_2$) 
different radion mass eigenvalues are obtained in different subregions, which contradicts the claim that there is one radion in the theory.
This inconsistency does not appear in the calculation of ref.~\cite{Lee:2021wau}.
Note that $k_1 = k_2$ corresponds to the zero tension case for the intermediate brane, $i.e.$ the intermediate brane is absent,
and the inequality $k_2 > k_1$ is required in order to avoid a tachyonic solution~\cite{Kogan:2001qx}. Since $E'=0$ at the intermediate brane implies $k_1 = k_2$, we conclude that it is an inconsistent condition.

Finally, let us describe our intuitive understanding of the existence of two radions in the three 3-brane setup.
For the two 3-brane setup, the radion degree of freedom can be considered as
the fluctuation of the IR brane compared to the UV brane
(see $e.g.$ appendix of ref.~\cite{Pilo:2000et}).
Now, if we introduce another brane (the intermediate brane),
there is also the fluctuation of the intermediate brane compared to the UV brane,
which corresponds to the second radion degree of freedom. 


	\bibliographystyle{utphys}
	\bibliography{rsc_bib}	
	
\end{document}